\documentclass[a4paper,fleqn,usenatbib,useAMS]{mnras}

\usepackage[T1]{fontenc}
\usepackage{ae,aecompl}

\usepackage{graphicx}
\usepackage{amsmath}
\usepackage{rotating}
\usepackage{pdflscape}
\usepackage{multirow}
\usepackage{amssymb}

\def\aap{{ A\&A}}

\def\aj{{AJ}}

\def\apj{{ApJ}}
\def\apjl{{ApJL}}

\def\mnras{{MNRAS}}

\def\pasa{{PASA}}

\def\apjs{{ApJS}}

\title[Observing simulated molecular clouds]
{What can simulated molecular clouds tell us about real molecular clouds?}

\author[A. Duarte-Cabral et al.]{A. Duarte-Cabral$^1$\thanks{E-mail: adc@astro.ex.ac.uk}, C. L. Dobbs$^1$ \\
$^1$ School of Physics, University of Exeter, Stocker Road, Exeter, EX4 4QL, U.K. \\
}

\date{Accepted 2016 February 25. Received 2016 February 25; in original form 2015 November 12}
\pubyear{2016}

\begin{document}
\label{firstpage}
\pagerange{\pageref{firstpage}--\pageref{lastpage}}
\maketitle

\begin{abstract}

We study the properties of giant molecular clouds (GMCs) from a smoothed particle hydrodynamics simulation of a portion of a spiral galaxy, modelled at high resolution, with robust representations of the physics of the interstellar medium. We examine the global molecular gas content of clouds, and investigate the effect of using CO or H$_{2}$ densities to define the GMCs. We find that CO can reliably trace the high-density H$_{2}$ gas, but misses less dense H$_{2}$ clouds. We also investigate the effect of using 3D CO densities versus CO emission with an observer's perspective, and find that CO-emission clouds trace well the peaks of the actual GMCs in 3D, but can miss the lower density molecular gas between density peaks which is often CO-dark. Thus the CO emission typically traces smaller clouds within larger GMC complexes. We also investigate the effect of the galactic environment (in particular the presence of spiral arms), on the distribution of GMC properties, and we find that the mean properties are similar between arm and inter-arm clouds, but the tails of some distributions are indicative of intrinsic differences in the environment. We find highly filamentary clouds (similar to the giant molecular filaments of our Galaxy) exclusively in the inter-arm region, formed by galactic shear. We also find that the most massive GMC complexes are located in the arm, and that as a consequence of more frequent cloud interactions/mergers in the arm, arm clouds are more sub-structured and have higher velocity dispersions than inter-arm clouds. 

\end{abstract}


\section{Introduction}

The true distribution of the different gas components of the interstellar medium (ISM) is quite complex, and one of the key problems is the fact that the ISM is a continuous medium. However, having some means of discretising the ISM is crucial to understand the properties of the different hierarchical structures that are formed by the gas (from giant molecular complexes, to small molecular clouds, clumps and ultimately star forming cores), as this hierarchy is essential for star formation to take place. Giant molecular clouds (GMCs), in particular, form the larger-scale reservoirs of molecular gas within which stars form. They have typical sizes of~$\sim$\,50\,pc, masses of~$\sim$\,10$^{3} - 10^{6}$\,M$_{\odot}$, temperatures of~$\sim10$\,K, and they are typically observed through surveys of CO emission lines as a tracer of molecular gas \citep[e.g.][]{Solomon1987,Dame01}. By studying such GMCs we have come to derive various relations that describe the global properties of molecular clouds and their ability to form stars \citep[e.g.][]{Schmidt1959,Kennicutt1998,Larson1981,Johnstone2004,Lada2010}. It is not clear, however, how the properties of GMCs may be affected by their galactic environment, and how that could affect their star formation.

Using CO as a tracer of GMCs (and a tracer of H$_{2}$ gas in general) yields a number of limitations, particularly because the lower density H$_{2}$ gas can be devoid of CO, or contain so little CO that the resulting CO emission is below observational sensitivities. Although CO may be a good tracer of the density peaks inside GMCs, there is thus an unknown amount of molecular gas not traceable with CO: the so-called CO-dark molecular gas \citep[e.g.][]{Klaassen2005}. In addition, observational studies of GMCs are complicated by a number of limitations, namely the inability to see the real 3D distribution of the gas, and instead using velocity information from molecular line emission as a proxy of the third spatial dimension. 
This is most critical for our Galaxy due to the line-of-sight confusion in the Galactic plane, and where the reliance on detailed kinematical models of the Milky Way \citep[e.g.][]{Reid2009} means determining kinematical distances is a common source of uncertainty. In nearby galaxies, line-of-sight confusion is less problematic, but studying GMCs has been limited by the resolution, where observations can identify individual GMCs, but are still short of resolving their inner substructure \citep[e.g.][]{Schinnerer2013}.

One important question regarding GMCs is whether they are essentially universal, or whether their properties depend on galactic environment, for example their passage through a spiral arm. Some studies have attempted to probe these environmental effects on GMC properties from observations of our Galaxy \citep[e.g.][]{Roman-Duval2010,Eden2012,Shetty2012} and nearby spiral galaxies \citep[e.g.][]{Hirota2011,DonovanMeyer2013,Colombo2014,Rebolledo2012,Rebolledo2015,Usero2015}. However, the results from these studies are still somewhat inconclusive, as some \citep[e.g.][]{Eden2012,DonovanMeyer2013} suggest that GMCs are insensitive to the physical conditions in their surroundings, while others have reported environment-dependent variations in GMC properties \citep[][]{Shetty2012,Colombo2014,Rebolledo2012,Rebolledo2015,Usero2015}. 

An alternative way to study GMCs, and their relation to the galactic environment, is through numerical simulations. This has the advantage that uncertainties regarding the conversion of CO to H$_{2}$, the relevance of dark CO, and distance ambiguities, can be tested. To date, there have only been few attempts to study GMCs in H$_2$ \citep[e.g.][]{DobbsBonnell2006,Nimori2013,Fujimoto2014,Khoperskov2015}. This is mainly because numerical models of galaxies often lack the resolution and/or many of the physical processes fundamental to capture the complexity of the ISM down to parsec scales. \citet[][]{Fujimoto2014} found that the mean global GMC properties are independent of environment, but the tails of the distributions do vary, although the lack of supernovae feedback in such simulations (which has a strong impact on the global distribution of gas) may limit the robustness of the results. Other numerical simulations have included feedback, but they typically just use density to define the clouds \citep[e.g.][]{Dobbs2015}, rather than CO emission. It is not clear how reliable, or how comparable the properties of such simulated GMCs are to observations.

The comparison between the 3D position-position-position (PPP) space of the simulations and the observable position-position-velocity (PPV) space of molecular line emission has been the focus of some studies in the literature. However, these are typically only on the scale of an individual cloud  \citep[e.g.][]{Shetty2010,Ward2012,Beaumont2013} or on galaxy-scales but with a face-on perspective \citep[i.e. equivalent to an extragalactic observer, e.g.][]{Pan2015}, and thus less severely affected by projection effects when compared to Galactic observations. Furthermore, most attempts to compare PPP and PPV perspectives do not model the emission with radiative transfer \citep[they typically assume a fixed H$_{2}$ density threshold above which we should have observable molecular clouds][]{Shetty2010,Ward2012,Pan2015}. As noted by \citet{Beaumont2013}, the topology of the CO emission can be decoupled from the morphology of the underlying H$_{2}$ gas, and this can influence the observable properties of molecular clouds.

In this paper, we study the distribution and properties of the molecular gas in a numerical simulation from \citet[][]{Dobbs2015}, capturing a fraction of a spiral arm, and inter-arm material moving towards the arm (see Figs.~\ref{fig:sim} and \ref{fig:top-down}), both in 3D space, and from an observer's perspective, by taking an edge-on perspective of a galactic disc, realistically mimicking the line-of-sight complications inherent to Galactic observations. This work includes a due treatment of the CO chemistry within the simulation as well as full non-local thermodynamic equilibrium (non-LTE) radiative transfer calculations to derive the observable CO emission. 
The simulation and the radiative transfer calculations are described in Section~\ref{section:method}. We investigate the effects of using CO as a tracer of H$_{2}$ gas, both in 3D space, and from an observer's perspective (PPV space), in Section~\ref{section:properties_gmc}. In Section~\ref{section:extreme_and_environment} we investigate the existence of any systematic correlation between the properties of clouds and their position with respect to spiral arms. We also examine a sub-sample of clouds which are particularly striking, and discuss our results in the context of linking the properties of GMCs to their larger scale environment. In Section~\ref{section:equilibrium}, we discuss the global equilibrium state of the GMCs, and finally, in Section~\ref{section:conclusions} we present our summary and conclusions.


\section{Method}
\label{section:method}

\subsection{The numerical model}
\label{section:model}

\begin{figure}
	\includegraphics[width=0.48\textwidth]{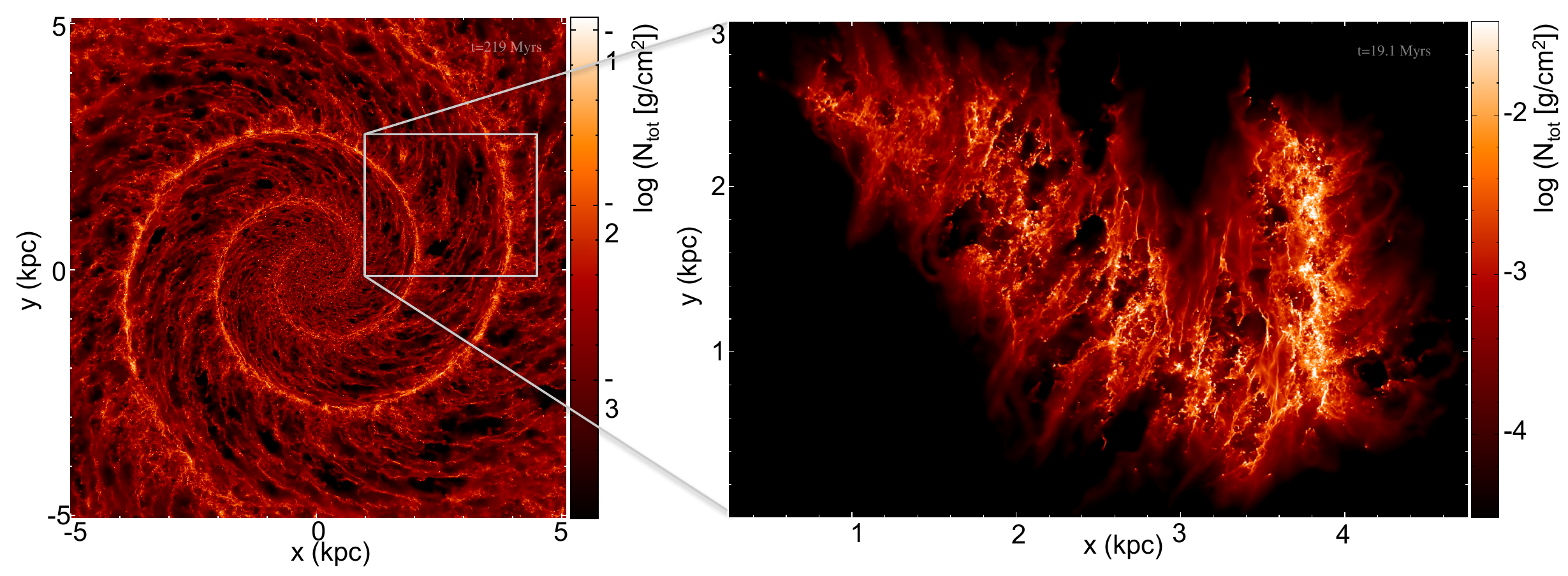}
	\vspace{-0.3cm}
	\caption{{\it Left:} Top-down total column density map of the galaxy model presented in \citet{Dobbs2013}. {\it Right:} Higher-resolution simulation of the portion of the galaxy-simulation lying within the box on the left panel \citep{Dobbs2015}.}
	\label{fig:sim}
\end{figure}

\begin{figure*}
	\includegraphics[width=0.90\textwidth]{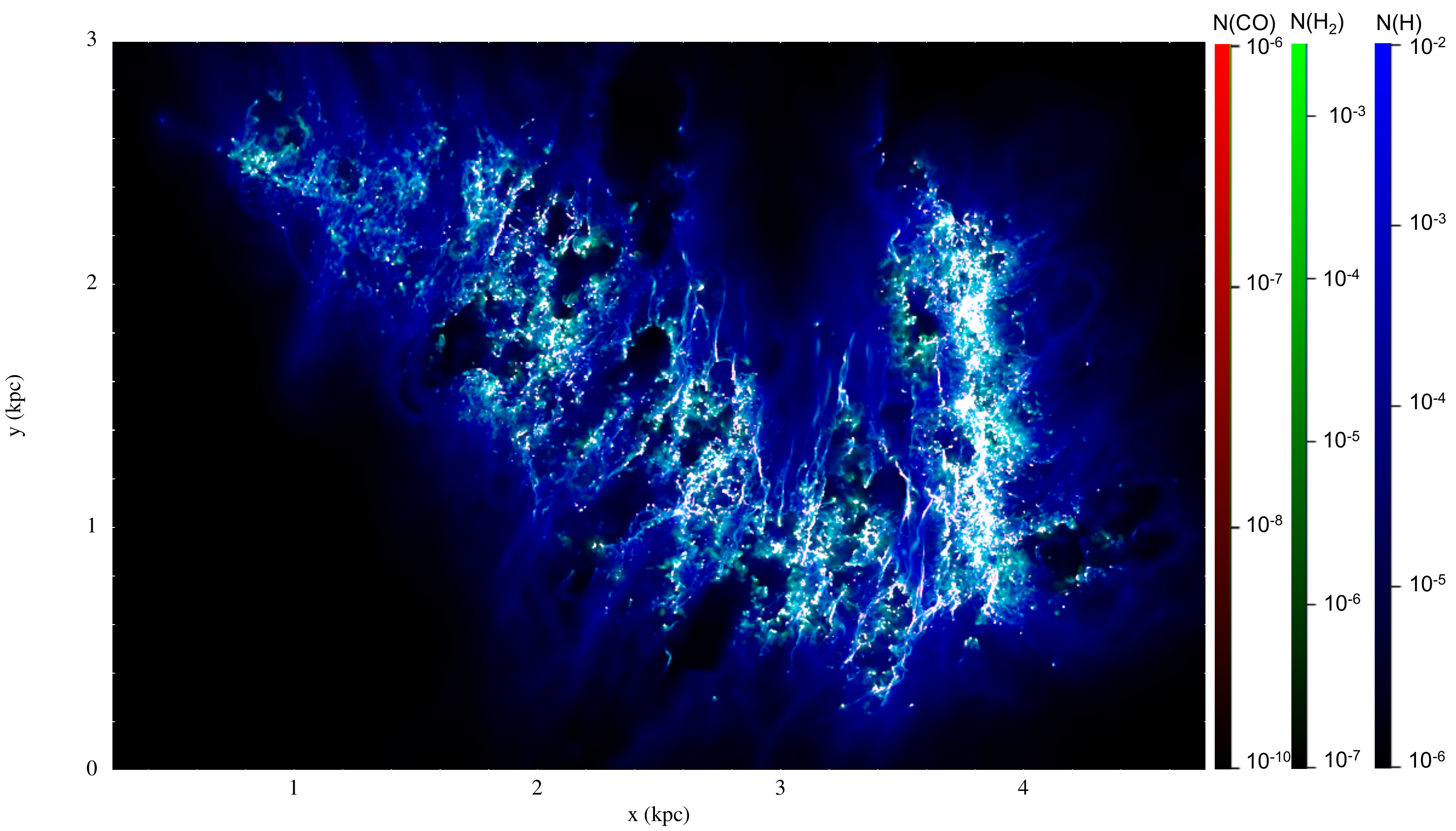}
	\caption{Top-down view of the simulation from \citet{Dobbs2015} used in this work, as a 3-colour (RGB) image of the column densities of CO (red), H$_{2}$ (green) and atomic H (blue), in units of g\,cm$^{-2}$. For the synthetic observations, we positioned the observer in the top-left corner, at (0,3,0)~kpc coordinates.}
	\label{fig:top-down}
\end{figure*}

In this paper, we study the population of clouds within the Smoothed Particle Hydrodynamics (SPH) simulation described in \citet[][]{Dobbs2015} at the timestep of 19.1\,Myr. This simulation is a section of the galaxy model presented in \citet{Dobbs2013} simulated at higher-resolution (see Fig.\,\ref{fig:sim}), so that we can properly resolve molecular clouds properties such as their morphology and dynamics (which was not the case with the full-galaxy model). This simulation has a particle mass of $\sim\,3.85$\,M$_{\odot}$, and it includes self-gravity, heating and cooling, and simple H$_{2}$ and CO formation as described in \citet{Dobbs08}, and \citet{Pettitt2014}. The minimum temperature of the gas in the simulation is 50\,K. This simulation also includes a two-armed spiral potential, as was used in the original simulation from \citet[][]{Dobbs2013}. Feedback is included using the same method as described in \citet{Dobbs11b}, where feedback is instantaneous, and inserted whenever gas lying above a 500\,cm$^{-3}$ density threshold is both bound and converging. Here we use the model with a feedback efficiency of $\epsilon=0.4$. Figure~\ref{fig:top-down} shows a top-down view of the column density of the different gas components of this simulation (CO in red, H$_{2}$ in green, and atomic H in blue), and full details about the simulation can be found in \citet[][]{Dobbs2015}.

Unlike in \citet[][]{DobbsPDC2015}, where the total gas densities were used to identify clouds, here we use either H$_{2}$ densities, CO densities, or CO emission. Moreover, we also choose to use an algorithm used by the observational community to find clouds (see Sect.~\ref{section:scimes}), so that our results can be readily compared with Galactic surveys (e.g. Schuller et al. in prep.). 
One reason for using a grid-based cloud extraction algorithm was so that the same code can be used to extract properties of clouds from a 3D position-position-position (PPP) space, and from synthetic observations in a position-position-velocity (PPV) space. 
To build the PPP datacubes, we extracted the densities from the SPH data onto a three-dimensional regularly spaced grid. We used gridcells of 5\,pc in size, and extracted the volume densities of H, H$_{2}$ and CO, from $x=0.25$ to 4.75\,kpc, $y=0$ to 3\,kpc, and $z=-0.4$ to 0.4\,kpc (see Fig.~\ref{fig:top-down}).

\subsection{Radiative transfer calculations (PPV space)}
\label{section:RT_calculations}

\begin{figure*}
\includegraphics[width=\textwidth]{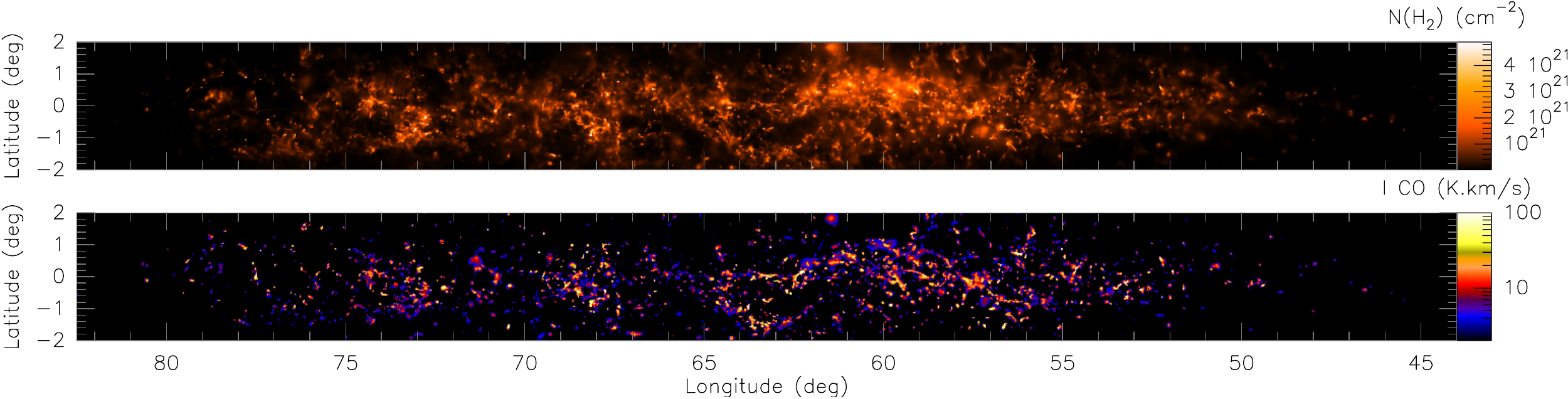}
\vspace{-0.3cm}
\caption{Synthetic observations of the simulation shown in Fig.\,\ref{fig:top-down}, created using the {\sc{torus}} radiative transfer code, with the observer positioned at (0,3,0)~kpc. The top panel shows the H$_{2}$ column densities, and the lower panel shows the integrated intensities of CO\,(1-0) emission, with a resolution of 36$''$, i.e. before convolution with a beam of 20 pixels FHWM.}
\label{fig:synth_obs}
\end{figure*}

We have used the {\sc{torus}} radiative transfer code \citep{Harries00} to post-process the SPH data described in Sect.~\ref{section:model}, and generate synthetic observations in galactic co-ordinates, similarly to \cite{duarte-cabral2015}. We did so by placing the observer inside the simulation, at the position $x=0$\,kpc, $y=3$\,kpc and $z=0$\,kpc, i.e. in the top-left corner of Fig.~\ref{fig:top-down} and in the galactic plane. This configuration is similar to observing the Galactic Perseus arm in the first quadrant, but the distances to the different features cannot be directly compared to the Milky Way due to the configuration of the model. 

The synthetic spectral cubes of CO\,($1-0$) emission (in PPV) are calculated using the molecular physics module of {\sc{torus}} as described in \cite{rundle_10}, which maps the SPH simulation onto an adaptive mesh refinement (AMR) grid. The CO\,($1-0$) datacube is generated without making the assumptions of local thermodynamic equilibrium (LTE) and the large velocity gradient (LVG) approximation \citep[e.g.][]{santander_2012}. This requires calculating non-LTE level populations of the CO molecule in each cell of the AMR grid. Once the level populations have been determined, the emissivity and opacity of each cell on the AMR grid can be calculated and a spectral cube of CO emission generated using a ray tracing method \citep[as in][]{douglas_2010,Acreman12}. The resulting spectral cubes have latitude-longitude-velocity co-ordinates, with velocity channels of 0.5\,km\,s$^{-1}$ (over a velocity range of $80$ to $200$~km/s), a pixel size of 36$''$, and are centred at a longitude of $l=60^{\circ}$. The integrated intensity map of the CO\,($1-0$) emission is shown in the lower panel of Fig.~\ref{fig:synth_obs} (with the column density map of H$_{2}$  also shown in the top panel). However, to make the PPV analysis more comparable to the PPP one (where the linear resolution is 10\,pc, from the 5\,pc grid cells), we have convolved the synthetic CO datacube with a Gaussian of 20 pixels full width at half maximum (FHWM). This effectively means that the resolution of the final synthetic datacube is of 0.2$^{\circ}$, which, for distances ranging from $\sim$1\,kpc to $\sim$4\,kpc, corresponds to a linear resolution of $\sim$3.5 to $\sim$14\,pc respectively. Note that even though we opt for an internal view of the galaxy, we do not assign a velocity to the observer and therefore the velocities are those with respect to rest, contrary to \cite{duarte-cabral2015} where we had taken the observer's velocity from its location in the galaxy. Even though this results in velocity values different to those we would observe in the Milky Way, it does not alter the relative  velocities between clouds (or the velocity gradients within clouds) which is what we are interested in.

We also generated an auxiliary synthetic spectral cube of CO\,($1-0$) emission (in PPV) from a top-down perspective, i.e. equivalent to an extragalactic observation of a face-on spiral galaxy. This external perspective of the emission was generated with a spatial resolution of 10\,pc (5\,pc grid cells), and using the less computationally-intensive assumption of LTE, that despite not providing the correct intensities, is sufficient for the purpose of retrieving the spatial distribution of the emission \citep[][]{duarte-cabral2015}.

\subsection{Cloud extraction method: {\sc scimes} algorithm}
\label{section:scimes}

In this paper, we have used {\sc scimes} \citep[Spectral Clustering for Interstellar Molecular Emission Segmentation, see][for full details]{Colombo15}, which is a code designed to identify GMCs in observations based on cluster analysis. While other available 3D cloud-extraction algorithms tend to segment the emission into individual emission peaks/clumps inside GMCs \citep[e.g. {\sc clumpfind}, or {\sc fellwalker} by][respectively]{Williams1994,Berry2015}, the advantage of {\sc scimes} is that  it is tailored to group different peaks together, making it more suitable to extract the larger complexes of clouds. In practice, this code considers the dendrogram tree of the 3D structures in the datacube \citep[as per the implementation of][to analyse astronomical datasets]{Rosolowsky2008} in the broader framework of graph theory, and groups different leaves together into ``clusters'' of leaves, based on some criteria (e.g. density, luminosity, and/or volume). For our particular usage of the code and given the resolution of our data, the dendrogram leaves correspond to peaks in density or intensity, typically what would be individual molecular clouds, while the ``clusters'' from {\sc scimes} correspond to GMCs (i.e. grouping of smaller clouds into larger cloud complexes). 

\begin{figure}
\includegraphics[width=0.5\textwidth]{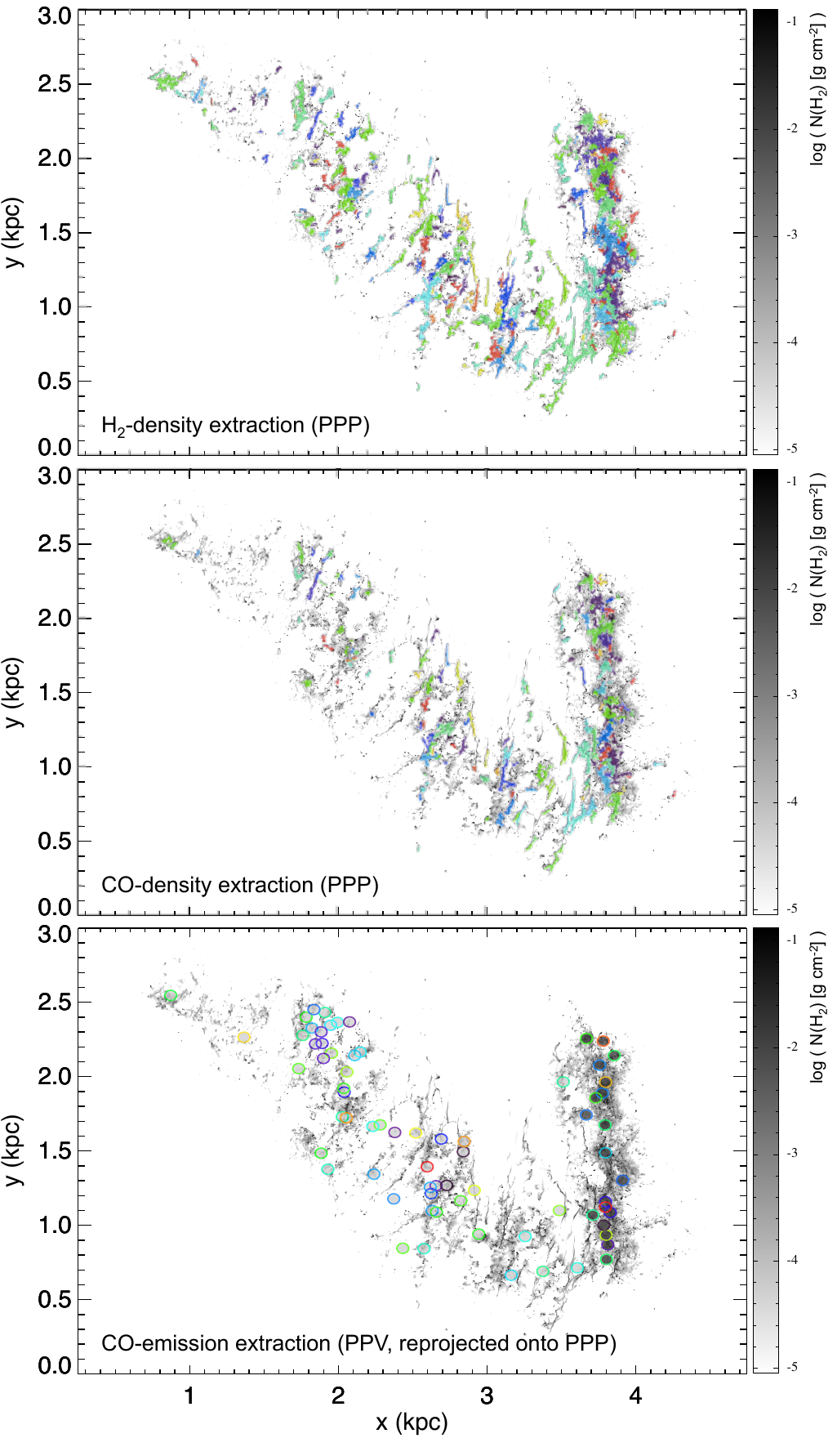}
\vspace{-0.4cm}
\caption{Top-down view of the molecular gas column density of the simulation in grey-scale, with the mask of the GMCs extracted with the {\sc scimes} code in colour, where each cloud is represented by one colour, which relates to the cloud $z$ coordinate (blue being lower $z$ values and red being higher $z$). The top panel shows the extraction of GMCs on the H$_{2}$ density (PPP) datacube, the middle panel shows the GMCs extracted from the CO density (PPP) datacube, and the lower panel shows the central position of the clouds extracted from the CO-emission (PPV), de-projected onto the PPP space. In all three cases we include the molecular cloud complexes from {\sc scimes} and any single leaf that was resolved.}
\label{fig:top-down-CO-molecular}
\label{fig:top-down-molecular}
\end{figure}

Even though both the dendrograms, and {\sc scimes} were originally designed to work on spectral datacubes (i.e. in PPV), they are built such that they are easily applicable to any 3D datacube. Therefore, we have used this code to find and extract GMCs from the H$_{2}$ and CO density datacubes (i.e. in PPP). We used both the density and the volume as a criteria for the clustering. This extraction provides an output datacube with a mask containing all the {\sc scimes} ``clusters'' found, a datacube with the mask of only the dendrogram leaves, and the entire catalogue of dendrogram structures. As we were also interested in the larger clouds that may simply have little substructure within them (the level of sub-structuring could be a consequence of both the grid and SPH resolution), we have also evaluated the single leaves in the dendrogram which had been excluded by the clustering algorithm. We then included in our final sample of clouds those single-leaves whose size was large enough to be well resolved (i.e. when the $2\sigma$ of the major axis of the structure was larger than 15\,pc). The results from the PPP extraction are summarised in Sect.~\ref{section:ppp}.

We also ran the extraction of GMCs on the synthetic CO (1-0) emission datacube (PPV), using the intensity and volume as the clustering criteria for  {\sc scimes}. Similarly to the PPP extraction, we also included all the dendrogram's single leafs which were well resolved, i.e. whose $2\sigma$ of the major axis was larger than 0.2\,$^{\circ}$ (20\,pixels). The results of this PPV extraction are detailed in Sect.~\ref{section:ppp_ppv}.

\subsection{SPH-based cloud extraction method}
\label{section:sph_extraction}

We briefly compare the extraction from {\sc scimes} with the SPH-based method used in \citet[][]{Dobbs2015} and \citet[][]{DobbsPDC2015}, where clouds are found using a friends of friends type algorithm. This algorithm first selects all particles above a given density, and then groups together particles within a certain distance ($l$) from each other into a single cloud. In \citet[][]{Dobbs2015}, this was done taking a threshold total density of 100 cm$^{-3}$ and a length $l$ of 2.5 pc. However, for a more direct comparison with the {\sc scimes} results, here we also investigate an SPH-based cloud extraction that adopts a minimum molecular (H$_{2}$) density of 10\,cm$^{-3}$, and a length $l$ of 5\,pc. The results from this comparison can be found in Sect.~\ref{section:scimes_vs_sph}.


\section{Cloud extraction results}
\label{section:properties_gmc}

\subsection{Density-based extraction (PPP)}
\label{section:ppp}

\subsubsection{GMC properties}
\label{section:ppp_properties}

From the 3D spatial cubes of H$_{2}$ and CO densities we extracted 350 and 195 resolved GMCs respectively, out of a total of 824 and 893 clouds. Figure~\ref{fig:top-down-molecular} shows the position and extent of all the resolved clouds from the H$_{2}$ densities on the top panel, and from the CO densities on the middle panel. For each of these resolved GMCs we derived a set of properties including: the distance from the spiral arm; the mass; the 3D major axis and aspect ratio, estimated by fitting the 3D density structure with an ellipsoid (the major axis was estimated as 2*$\sigma_{major}$ of the ellipsoidal fit, and the aspect ratio as the major axis divided by the average of the two other axis); the line of sight velocity ($v$) and velocity dispersion ($\sigma_{v}$), by assuming the observer to be at the same position as for the synthetic observations; and the level of sub-structuring (i.e. the number of leaves within each GMC, as per the dendrogram tree). For clouds extracted from the CO densities, we then need to convert the CO masses into H$_{2}$ masses. To do so, we estimated the ratio between CO and H$_{2}$ densities in the datacube (in units of g\,cm$^{-3}$), for all the pixels where CO clouds had been extracted, and found a mean value of $\sim7\times10^{-4}$. In terms of number density, this corresponds to a mean CO abundance with respect to H$_{2}$ of $\sim0.5\times10^{-4}$, marginally lower than the fiducial value typically used in the Galaxy (of $\sim1\times10^{-4}$), although it can reach up to $\sim2\times10^{-4}$ in the denser regions. The distributions of the different properties of GMCs can be seen in the histograms of Fig.~\ref{fig:histograms_ppp}, and the statistical properties are summarised in Table~\ref{tab:stat_properties}, including the median values and their deviation (using the first and third quartiles), as well as the mean, standard deviation, skewness $S$, and kurtosis $K$ of the distributions. Both the skewness and kurtosis are reflective of the profile of the distributions, that can be seen in the histograms of Fig.~\ref{fig:histograms_ppp}. In particular, skewness is a measure of the symmetry of a distribution: a symmetric distribution has a skewness value close to zero, while larger positive and negative skewness values correspond to asymmetric distributions with tails towards higher or lower values around the mean, respectively. Kurtosis on the other hand, is a measure of how peaked a distribution is compared to a Gaussian distribution: a kurtosis of zero corresponds to a Gaussian profile, a negative kurtosis reflects flatter distributions, and positive kurtosis corresponds to more strongly peaked distributions.

\begin{table*}
\caption{Statistical properties of GMCs from the H$_{2}$ and CO 3D PPP extractions, and from the CO emission PPV extraction.}
\label{tab:stat_properties}
\renewcommand{\footnoterule}{}  
\begin{tabular}{l c c c c l c c c c l c c c c}
\hline 
\hline
\multirow{2}{2.0cm}{Property}	& \multicolumn{4}{c}{\it GMCs from H$_{2}$ densities}	&  & \multicolumn{4}{c}{\it GMCs from CO densities} & & \multicolumn{4}{c}{\it GMCs from CO (1-0) emission}		\\
						& \scriptsize{Median$^{(a)}$} & \scriptsize{Mean$^{(b)}$}  & \scriptsize{$S$} & \scriptsize{$K$} & & \scriptsize{Median$^{(a)}$} & \scriptsize{Mean$^{(b)}$}  & \scriptsize{$S$} & \scriptsize{$K$} & & \scriptsize{Median$^{(a)}$} & \scriptsize{Mean$^{(b)}$}  & \scriptsize{$S$} & \scriptsize{$K$}\\
\hline 
\hline
$log(M$ [M$_{\odot}$]$)$		& $3.4\pm0.8$	& $3.5\pm1.0$	& 0.4 & -0.7	& & $4.0\pm0.6$	& $3.9\pm1.1$	& -1.5 & 4.5	& & $3.5\pm0.5$	& $3.7\pm0.7$	& 0.7 & -0.33	\\
\hline
Aspect ratio				& $3.3\pm1.1$ 	& $3.9\pm1.9$  & 1.5 & 2.7 	& & $4.1\pm1.8$	& $4.9\pm2.7$	& 1.1	 & 0.9 	& & $1.9\pm0.4$	& $2.1\pm0.7$	& 1.3	 & 2.0	\\
\hline
Major axis [pc]			& $27\pm9$ 	& $36\pm27$	& 3.5	 & 17.3	& & $26\pm10$		& $34\pm24$	& 2.0	 & 4.5	& & $18\pm8$		& $28\pm29$	& 2.6	 & 6.7	\\
\hline
$\sigma_{v}$ [km\,s$^{-1}$] 	& $2.1\pm1.0$ & $2.6\pm2.1$ 	& 3.2 & 16.0 	& & $2.0\pm0.9$	& $2.2\pm1.6$	& 2.3 & 8.6	& & $3.1\pm0.8$	& $3.9\pm2.4$	& 1.9 & 3.3 	\\
\hline
Nb. of leaves 				& $1.0\pm1.0$	& $3.5\pm5.8$	& 5.3	 & 35.5	& & $3.0\pm1.0$	& $3.3\pm3.1$	& 2.4 & 8.2	& & $3.0\pm1.0$	& $3.4\pm3.9$	& 2.0 & 3.3	\\
\hline 

\end{tabular}
\flushleft
$^{(a)}$ Uncertainties derived as the mean value of the absolute deviation of the first and third quartiles from the median.\\
$^{(b)}$ Quoted uncertainty refers to the standard deviation.\\
\end{table*}

We also calculated the ratio between atomic and molecular mass, $M($H$)/M($H$_{2})$, for all the resolved GMCs, and found an average value of $\sim5.4$, reaching $\sim2.5$ in the densest regions (i.e. only up to $\sim30\%$ molecular). Although this represents a smaller molecular fraction than what one could expect for ``molecular'' clouds, the density threshold in the simulation is relatively low, and hence we do not capture the ``purely-molecular'' zones within the GMCs. A similar value (of $\sim5$) is found for the total $M($H$)/M($H$_{2})$ in the simulation, which is a significant improvement with respect to the lower resolution galaxy-simulations \citep[with $M($H$)/M($H$_{2})$ values reaching as high as $\sim$80, e.g.][]{duarte-cabral2015}, and more in-line with observations of the Milky Way with $M($H$)/M($H$_{2})$ of $\sim6$.

\subsubsection{H$_{2}$ PPP $vs$ CO PPP}
\label{section:ppp_h2_c2}

Since CO is often used as a proxy for tracing the molecular gas content of clouds, here we explore this assumption, by comparing the GMCs extracted from the CO densities with the clouds from H$_{2}$ densities. In Fig.~\ref{fig:top-down-molecular} we compare the position and extent of all the resolved clouds from the H$_{2}$ and CO densities, and we see that typically CO clouds are more compact, and confined to the higher H$_{2}$ density regions (see also Fig.~\ref{fig:top-down}). 

\begin{figure*}
\centering
\includegraphics[width=0.7\textwidth]{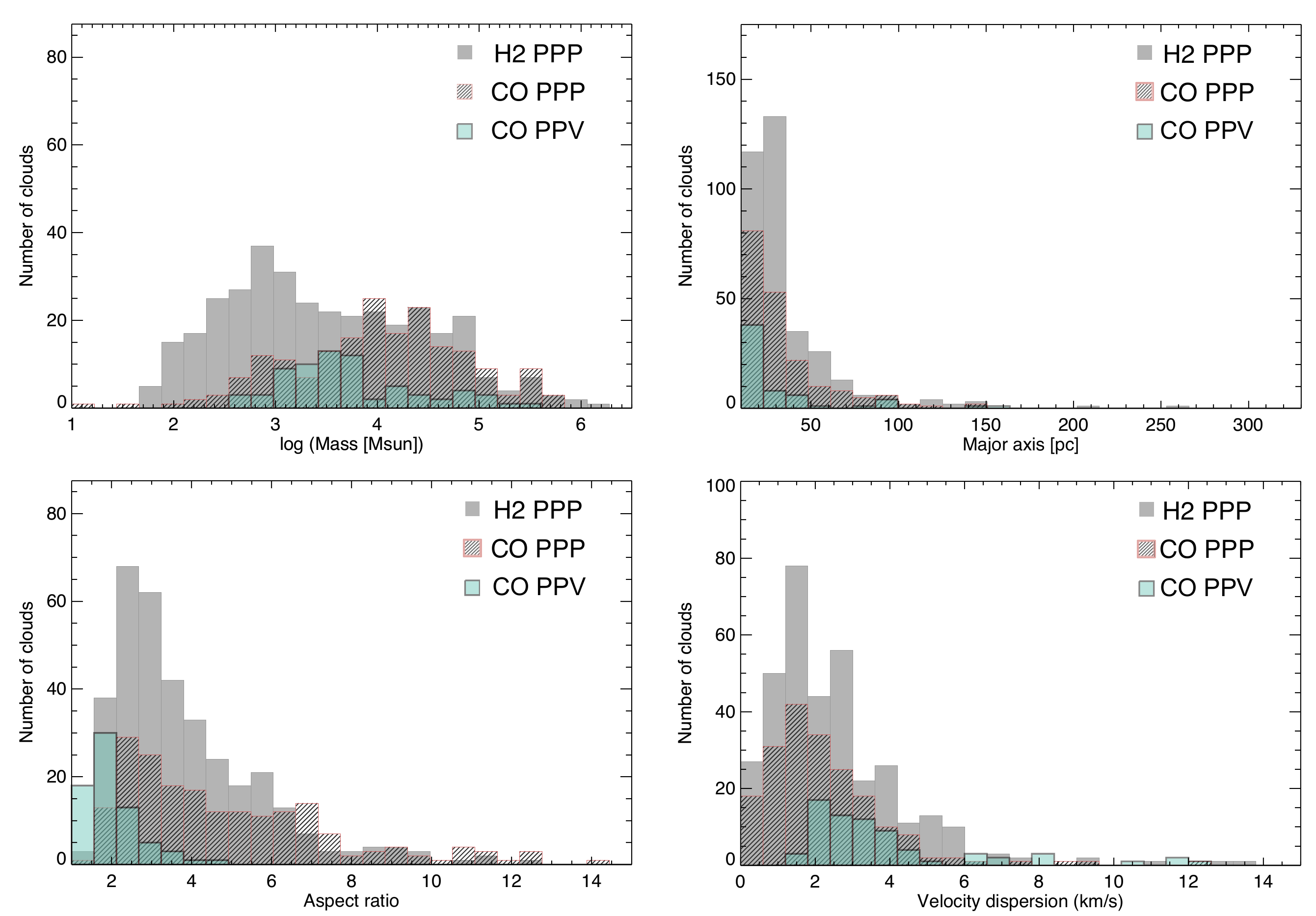}
\caption{Histograms of the properties of clouds extracted from the PPP cube with H$_{2}$ densities (grey-filled histograms), and CO densities (black-dashed histograms); as well the properties of the clouds extracted from the PPV cube of CO emission (light-green histograms). The properties are: mass (top-left), aspect ratio (bottom-left), the major axis (top-right), and velocity dispersion (bottom-right). For the PPP cubes, the aspect ratio and major axis are estimated in 3D, while for the PPV cube these are projected sizes (onto plane of the sky). The masses for the CO PPV clouds are calculated using a X$_{\rm CO}$ factor. Note that the CO emission cube was created with TORUS assuming a turbulent velocity of 1\,km\,s$^{-1}$, and a velocity channel of 0.5\,km\,s$^{-1}$, which means that we cannot resolve velocity dispersions below 1\,km\,s$^{-1}$.}
\label{fig:histograms_ppp}
\label{fig:histograms_ppp_ppv}
\end{figure*}

Figure~\ref{fig:histograms_ppp} shows the comparison of the distributions of cloud properties from these two extractions (in grey and black-dashed histograms). From that and Table~\ref{tab:stat_properties}, we can see that the distribution of cloud properties as retrieved from the H$_{2}$ or the CO PPP extraction are similar, even though the CO extraction picks up a smaller number of clouds, mainly tracing the denser H$_{2}$ clouds.  
The less massive, and smaller clouds are increasingly missed with CO (see top panels of Fig.~\ref{fig:histograms_ppp}),  due to CO being 
less abundant in the lower density regions, and/or simply being unresolved in the smaller H$_{2}$ clouds. Because CO is only tracing the higher density regions (best seen in Fig.~\ref{fig:top-down-molecular}) some of the larger H$_{2}$ clouds of the sample are also split into several smaller clouds in CO.
Despite that, the GMCs from CO densities have in fact higher aspect ratios, indicating more elongated clouds (see lower-left panel of Fig.~\ref{fig:histograms_ppp}). While this might be surprising, given that the clouds are more ``broken-up'' with the CO density extraction, the highly filamentary clouds are enhanced with the CO densities, as we are most sensitive to the high density ridge within filaments, and less sensitive to a lower density ``envelope'' of H$_{2}$, or the including of nearby ``satellite'' clouds, that would decrease the aspect ratio. In terms of velocity dispersion, the two distributions are very similar, with the CO density extraction missing a similar number of clouds throughout the distribution (see lower-right panel of Fig.~\ref{fig:histograms_ppp}). 

\subsubsection{{\sc scimes} PPP $vs$ SPH-based method}
\label{section:scimes_vs_sph}

The SPH-based algorithm is able to resolve the higher density regions to smaller spatial scales than those we can recover from a fixed size grid of 5\,pc. Therefore both SPH-based extractions (using total or H$_{2}$ density thresholds) typically fragment clouds more than the grid-based extraction (clouds in the SPH-based extraction are more similar to the leaf-level of the dendrograms). We can see this by comparing the mass distribution of clouds, as per the top panel of Fig.~\ref{fig:histograms_ppp}, with Fig.~8 of \citet{Dobbs2015} which used the SPH-based method on total densities. The SPH-based extraction recovers clouds with total masses (of H\,$+$\,H$_{2}$) up to 10$^{5}$\,M$_{\odot}$, with a steep mass spectrum, while {\sc scimes} recovers clouds with H$_{2}$ masses up to 10$^{6}$\,M$_{\odot}$, with a flatter mass spectrum. Nevertheless, most clouds are well recovered with both algorithms, and in particular the longer filaments of the simulation are recovered similarly in both extractions.

\subsection{CO emission-based extraction (PPV)}
\label{section:ppp_ppv}

While the previous section described properties of GMCs as per the actual 3D distribution of the molecular gas, observers cannot do the same exercise, as observations are limited to a 2D space plus a line-of-sight velocity of the gas. For Galactic observations in particular, observers often use the third dimension of spectral velocity to give some indication about the third spatial dimension, using our knowledge of the rotation curve of the Galaxy. In this section we test if, by taking the perspective of a Galactic observer, i.e. in PPV, the properties of the clouds as derived from the PPP cubes can still be recovered.

\subsubsection{GMC properties}
\label{section:ppv_properties}

Using {\sc scimes} on the synthetic CO$(1-0)$ PPV datacube, we extracted a total of 203 clouds, but only 71 of those are well resolved (i.e. have a 1$\sigma$ major axis greater than 10\,pixels). In order to estimate the physical properties of these PPV clouds, similarly to Sect.~\ref{section:ppp_properties}, we require information on the distance. Typically, observers rely on techniques such as parallax, or kinematical distances (which in turn rely on a kinematical model of the Galaxy) to derive distances to individual clouds. Although we could potentially use the line of sight velocities to determine the kinematical distances of clouds, we will refrain from doing so because it could affect our results as it would not allow for relative motions of clouds to be disentangled from the galactic motions. Instead, we have opted to use our knowledge of the 3D (PPP) physical position of clouds to determine their distances. We did so by computing, for each PPP cell, its correspondent PPV position and distance from the observer, and effectively built a PPV datacube containing the distances. However, although we can easily know where each grid of the PPP physical space should be on the PPV (angular) space, the opposite direction is not trivial, because each PPV gridcell often includes a combined contribution from several cells of the PPP space. To circumvent this problem, whenever several PPP cells contributed to a given PPV cell, we keep only the distance of the PPP cell with the highest CO density, under the assumption that this PPP cell is likely the one that will contribute to most of the CO emission. We then used this distance datacube to estimate the mean distance to any cloud detected in the PPV space, calculate their masses and physical (projected) sizes, and de-projected the central position of each cloud in the PPV back onto the original PPP space, as shown in Fig.~\ref{fig:top-down-CO-molecular} (lower panel). 

\begin{figure*}
\includegraphics[width=0.95\textwidth]{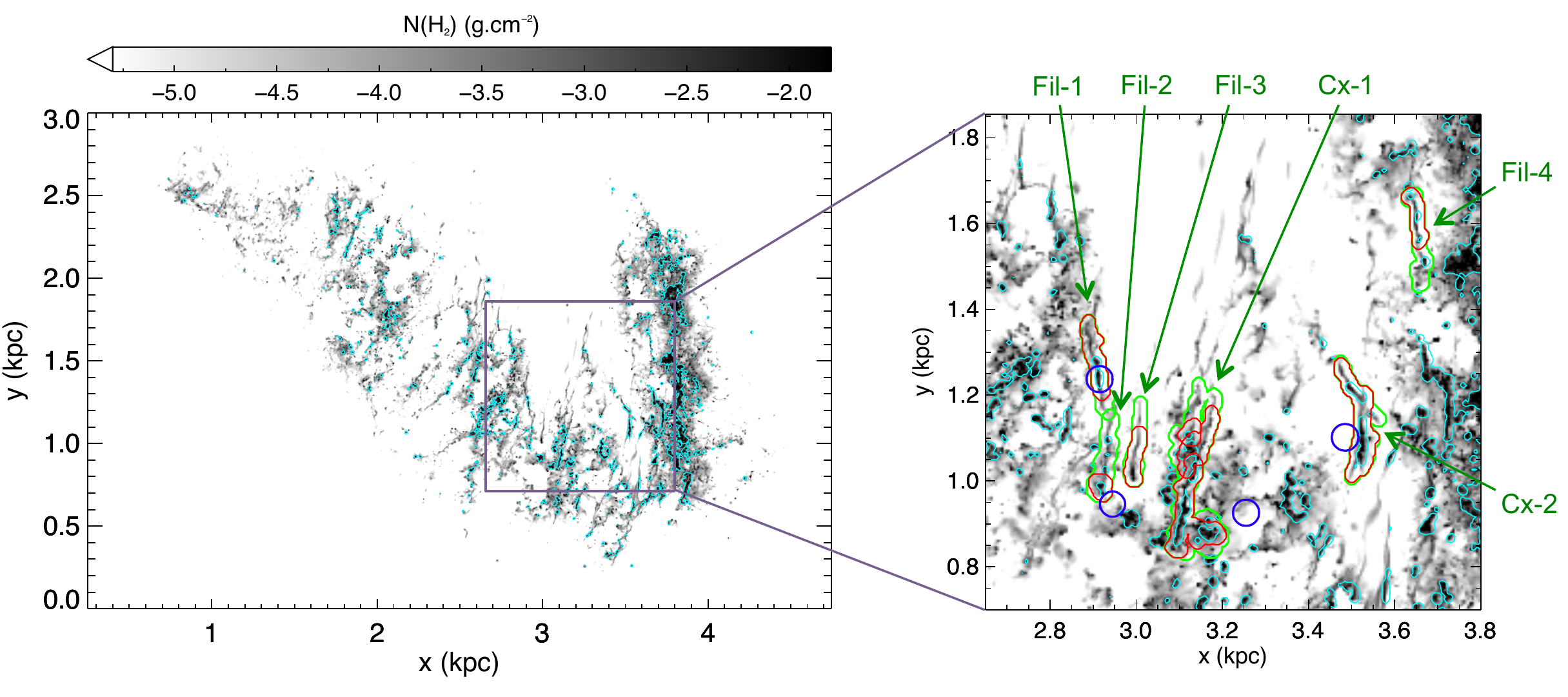}
\vspace{-0.3cm}
\caption{ Top-down view of the molecular gas column density of the simulation in grey-scale, with the turquoise contours showing the top-down CO (1-0) integrated intensity at a level of 1\,K\,km\,s${-1}$, mimicking an extragalactic perspective, with a resolution matching that of the underlying column density map. The right panel shows a zoom in of a portion of the simulation, with six of the extracted H$_{2}$ PPP clouds highlighted in green contours, and labeled as Fil-1 to 4 (for filamentary clouds) and Cx-1 to 2 (for complex clouds). For these clouds, we show the corresponding CO PPP clouds in red contours, and the central position of the matching CO PPV clouds de-projected onto the PPP space as dark blue circles.}
\label{fig:top-down-emission}
\end{figure*}

To estimate the H$_{2}$ cloud masses from the CO (1-0) integrated intensities, we have used the Galactic X$_{\rm CO}$ conversion factor of $2\times10^{20}$\,cm$^{-2}$\,K$^{-1}$\,km$^{-1}$\,s \citep[e.g.][]{Dame01}, which we confirmed to be a good conversion factor for these simulations as well (using the auxiliary synthetic H$_{2}$ column density map). Finally, the estimated sizes and elongations are 2D quantities in this case, as projected onto the plane of the sky. The results from this extraction are summarised in Table~\ref{tab:stat_properties} (right columns), and in Fig.~\ref{fig:histograms_ppp_ppv} (blue histograms). From here we can see that the typical masses of clouds we retrieve from the CO emission are smaller than the CO-density clouds, and the aspect ratios are lower. This is not totally unexpected, as the CO-bright clouds are only tracing a fraction of the clouds that we see in the CO-density cube, which are again just tracing a fraction of the entire molecular clouds as seen in H$_{2}$. 

This is better seen in Fig.~\ref{fig:top-down-emission}, where we highlight the distribution of CO emission from an extragalactic perspective, where line-of-sight confusion is less severe than for Galactic observations. The regions with CO emission are shown with turquoise contours, which display a sparsity of CO emission with respect to the distribution of H$_{2}$ gas (in grey-scale). The right-hand panel of Fig.~\ref{fig:top-down-emission} zooms in onto a few examples of clouds extracted from the H$_{2}$ densities (as green contours), with their corresponding resolved CO-density clouds (in red). We can see that the CO-PPP clouds are typically smaller portions of bigger H$_{2}$ clouds, or even break big complexes into several separate clouds (e.g. the cloud labeled as Cx-1 is split into three separate CO-density clouds). If we now look at the turquoise contours outlining the CO-emission, we can clearly see it only traces the very peaks of density, often unresolved. In fact, even relatively strong density peaks can be devoid of CO emission (see e.g. the cloud labelled as Fil-3). This is something that one should bear in mind when analysing the properties of molecular clouds. The observable molecular clouds are simply the higher density peaks of much larger complexes of molecular gas, and are far from being isolated structures, de-coupled from their surroundings.

Aside from the fact that the CO emission does not trace the entire underlying molecular gas structure, some of the most severe factors that can affect the correct identification of molecular gas structures and thus their statistical properties, are in fact the combination of perspective and resolution. We explore these limitations in the following section.

\subsubsection{Match between PPP and PPV GMCs}
\label{section:ppv_match}

\begin{figure}
\includegraphics[width=0.48\textwidth]{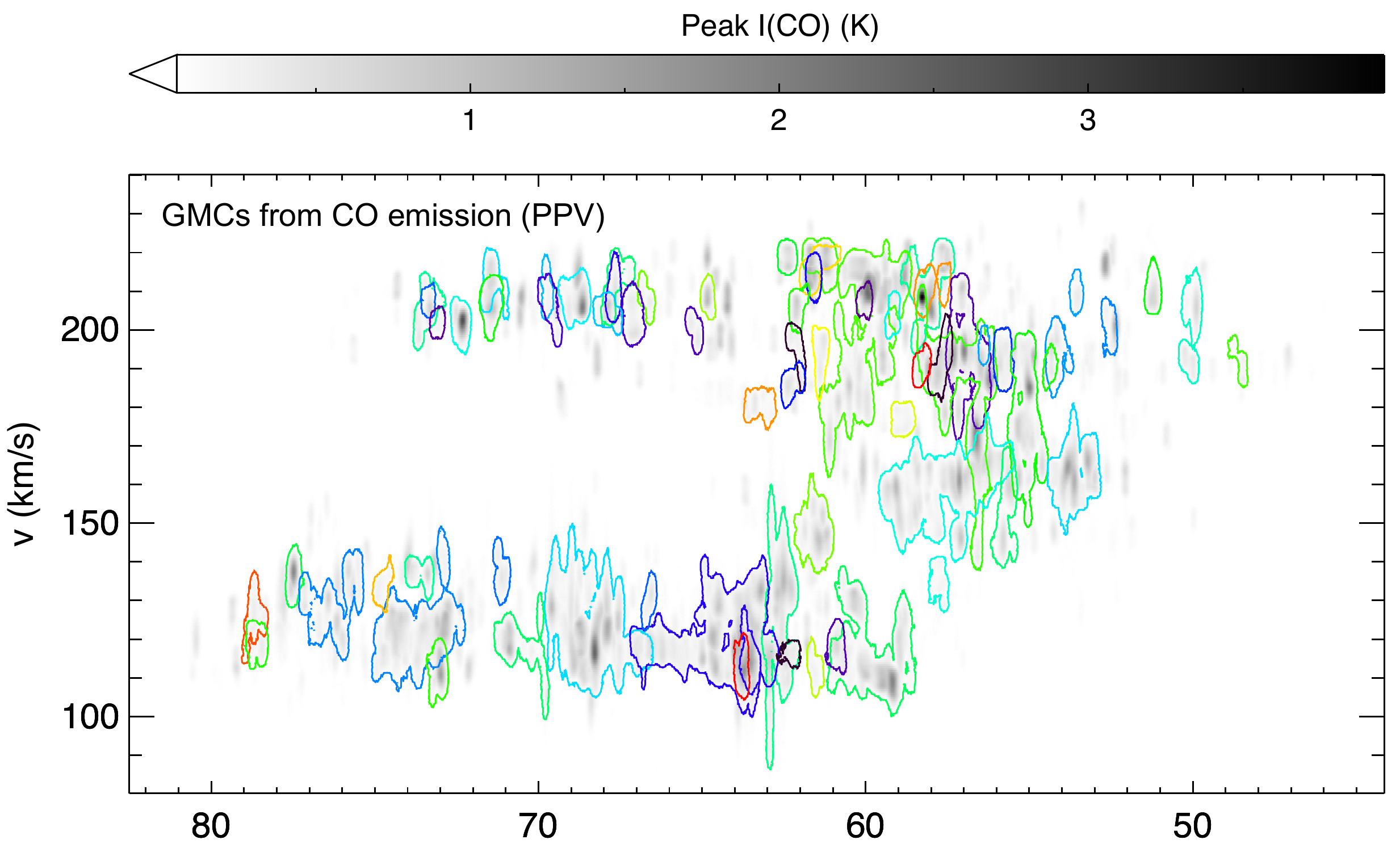}
\includegraphics[width=0.48\textwidth]{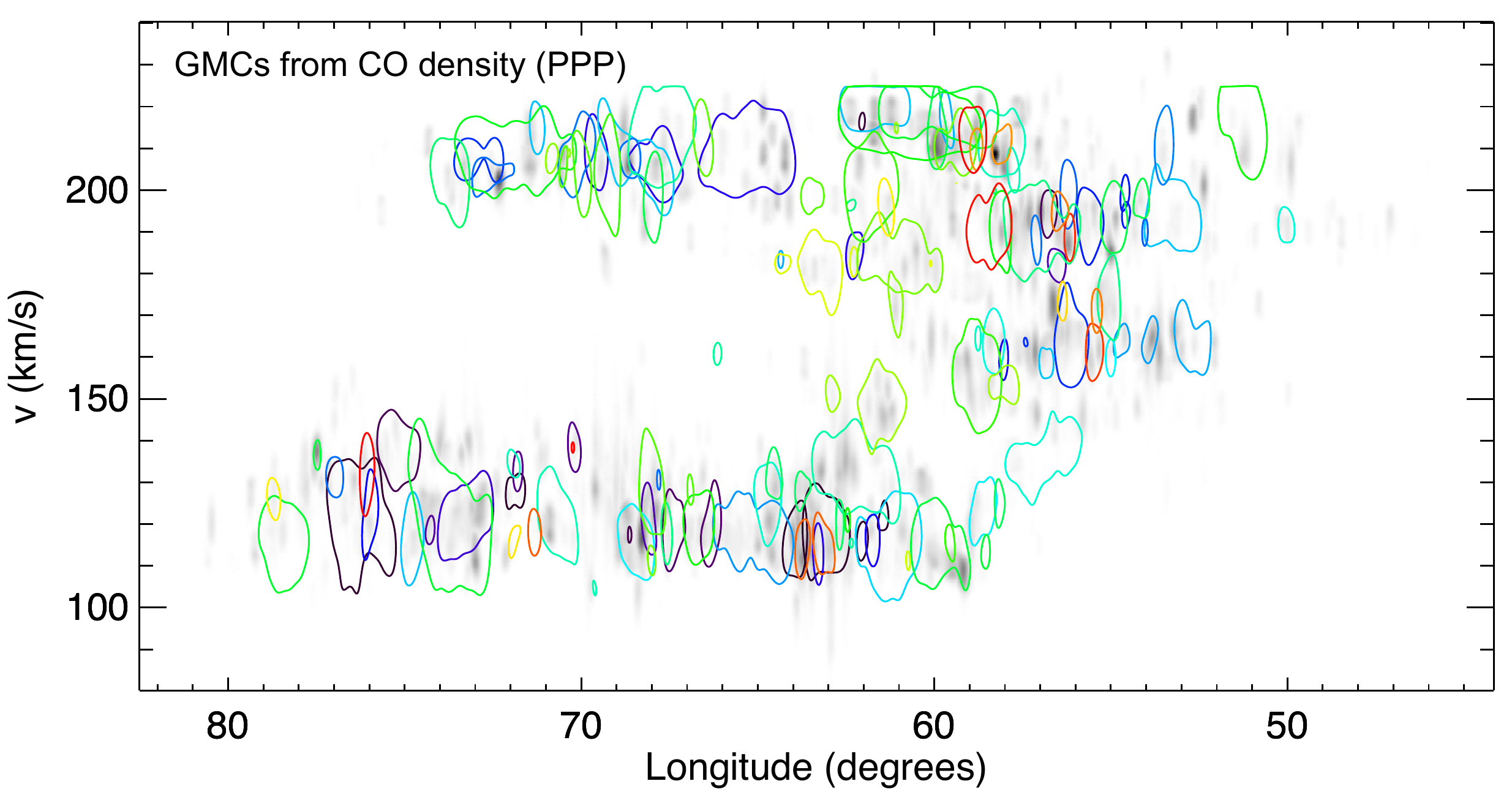}
\vspace{-0.3cm}
\caption{Top: Longitude-velocity plot with the peak CO intensity in grey-scale, and the position of the clouds extracted from the CO emission datacube in contours, with colour representing the $z$ coordinate of the clouds in the PPP (as in Fig.\,\ref{fig:top-down-molecular}), including only the ``clusters'' from {\sc scimes} and well resolved single leaves. Bottom: Same as top panel, but now the contours refer to the clouds extracted from the CO densities in PPP, projected into PPV, for comparison.}
\label{fig:CO_PPV}
\end{figure}

Clouds in PPV can suffer from strong blending, particularly with an edge-on perspective of the galactic disc observed at a relatively coarse angular resolution, where different clouds at different distances are easily merged onto the same point in PPV. Indeed, in the right-hand panel of Fig.~\ref{fig:top-down-emission}, we show the de-projected positions of the CO-PPV clouds (in blue circles) that correspond to the contoured H$_{2}$-PPP clouds (in green). In most cases, the distance we derive for the PPV cloud is not placing the cloud exactly within its PPP counterpart, simply because the PPV cloud contains several (merged) PPP clouds, and therefore the derived distance is an average among all the clouds included. As a consequence, the projected extent of the CO-PPV clouds do not match that of the original PPP clouds (see Sect.~\ref{section:extreme} and Appendix~\ref{app1}), suggesting that the derived statistical properties for the CO-PPV clouds are not representative of the actual underlying population of clouds.

In addition, given the large range of distances covered by galactic plane observations, the fixed angular resolution of observations means that the physical scales that we can resolve vary significantly along the line of sight, which can introduce further biases on the overall distribution of cloud properties. To better visualise this effect, Fig.~\ref{fig:CO_PPV} shows the clouds from the two extractions (CO emission clouds on the top panel, and CO density clouds in lower panel) as coloured-contours in a longitude-velocity plot, with the peak CO emission shown in greyscale. This figure is particularly useful to compare the spatial and spectral extent of the clouds from the two extractions. In particular, it shows that for nearby clouds (with velocities of $\sim$200\,km\,s$^{-1}$), the CO-density clouds are typically larger in angular size than the CO-emission clouds, while the inverse happens for the most distant clouds (at velocities of $\sim$120\,km\,s$^{-1}$). This is because nearby clouds are better resolved in the CO emission cube than in our PPP grid (hence PPP clouds are broken up into several individual CO-PPV clouds), while the more distant clouds (in the spiral arm) are better resolved in the PPP grid than in the CO emission cubes (hence several PPP clouds are grouped together into one physically large cloud in PPV). 

Given that the resolution element in these cubes is close to the typical size of the GMCs (i.e. $\sim10$\,pc resolution for clouds of $\sim50$\,pc in size), we explored how this blending problem would change for higher physical/angular resolution cubes (see App.~\ref{app1}). We found that blending becomes significantly less severe when we can properly resolve the distribution of the gas within the GMCs, down to sub-parsec scales (i.e. resolving scales at least two orders of magnitude below the size of the cloud). By doing so, line-of-sight confusion is minimised, and the only remaining issue then becomes the lack of CO at lower-densities, which is more easily dealt with (Sect.~\ref{section:co_bright}).

In summary, from our analysis of PPP clouds seen from an observer's perspective in PPV, we find that the most limiting factor is the angular/spatial resolution of the data, as this is what will dictate the ability to distinguish different clouds as separate emission peaks. This is in agreement with the results from \citet[][]{Beaumont2013}, where they conclude that in regions where the filling factor of emitting material is large, the effects of superposition and confusion on the derived properties of clouds can be severe. Here we find that in order to avoid line-of-sight confusion and properly probe large molecular cloud complexes of tens-of-parsec size-scales (distributed over a large range of distances), it is essential to resolve down to sub-parsec physical scales (see Sect.~\ref{section:extreme}).

\begin{figure}
\flushright
\includegraphics[width=0.46\textwidth]{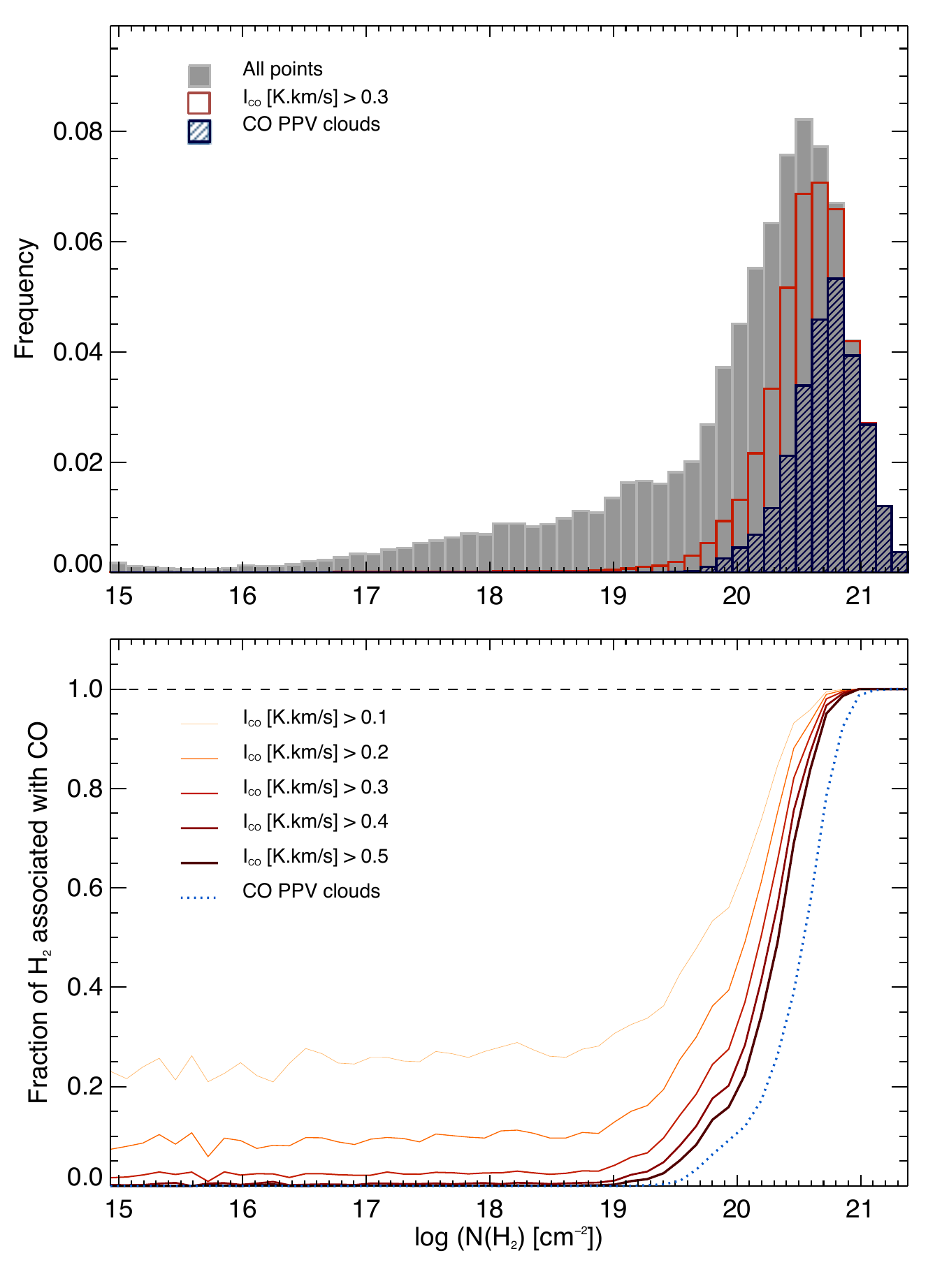}
\vspace{-0.3cm}
\caption{Distribution of H$_{2}$ column densities and their association with CO emission. On the top panel, the grey histogram shows the distribution of the H$_{2}$ column densities in the entire synthetic data (as per Fig.~\ref{fig:synth_obs}, convolved to the 0.2$^{\circ}$ resolution); and the coloured histograms show the distributions of H$_{2}$ column densities for regions where the integrated CO (1-0) emission is higher than 0.3\,K\,km\,s$^{-1}$ (in orange), and where a CO cloud has been extracted with the {\sc scimes} code (blue-dashed histogram). This shows that CO is well associated with all gas with H$_{2}$ column densities above $\sim5\times10^{20}$\,cm$^{-2}$, even though it can trace some fraction of the gas with H$_{2}$ column densities of a few 10$^{19}$\,cm$^{-2}$. The bottom panel, shows the fraction of pixels for a given H$_{2}$ column density beam, which have some CO emission associated. The thin-orange to thick-red lines correspond to different thresholds for the CO emission (mimicking different observational noise levels), and the blue dotted line corresponds to all extracted CO PPV clouds.}
\label{fig:CO_bright_threshold}
\end{figure}

\subsubsection{CO-bright threshold}
\label{section:co_bright}

As mentioned in the previous sections, one of the main problems we found for the CO emission clouds is that they are typically tracing smaller portions of larger GMC complexes. Although this may seem discouraging from an observer's point of view, the combined information from both H$_{2}$ column densities (e.g. from dust continuum emission) and CO emission should prove to be enough to re-gain access to the larger complexes of molecular clouds, even if a portion of these is still CO-dark. For example in Fig.~\ref{fig:synth_obs} it is clear that the H$_{2}$ column densities are capable of tracing the lower density regions that connect the otherwise separate peaks of CO emission. The reason for this CO-dark gas is one of two possibilities: either there is a regime where there is molecular gas but CO is not yet formed (e.g. if C is still in the form of atomic carbon, and not carbon monoxide); or, although there is a normal abundance of CO with respect to H$_{2}$, the column of CO is too small for it to be bright enough to be observed. Indeed the CO emission becomes very quickly very weak towards the lower column densities, and this can impose a strong limitation on the ability to trace molecular gas at low column densities - not because CO is not there, but because it is not bright enough \citep[e.g.][]{Burton2015}. 

To investigate whether there is a particular threshold above which the molecular gas is well traced by CO-bright emission, we compare the CO PPV clouds with the H$_{2}$ column density maps (from a Galactic perspective). In the top panel of Fig.~\ref{fig:CO_bright_threshold}, we show the distribution of H$_{2}$ column densities for all pixels in the synthetic data (in grey), as well as for the regions where CO clouds have been extracted (in blue) and within regions where the CO (1-0) integrated intensities were above 0.3\,K\,km\,s$^{-1}$ (in orange). We can see that high column densities always have associated CO emission (and CO clouds). But as expected, at lower column densities, CO becomes less bright, and eventually ceases to be detectable. Therefore, there is a significant fraction of lower column density pixels that are CO-dark, i.e. that have little or no CO emission associated. This is better illustrated in the lower panel of the same Fig.~\ref{fig:CO_bright_threshold}, where we plot the fraction of pixels within each H$_{2}$ column density bin that are associated with CO emission. Whilst 90-100$\%$ of pixels above a column density of $\sim5\times10^{20}$\,cm$^{-2}$ have significant CO emission, it rapidly decreases below that, and only 10$\%$ of the pixels with an H$_{2}$ column density of 10$^{20}$\,cm$^{-2}$ are effectively associated with the CO PPV clouds we extracted. Even if considering the weakest emission, e.g. as deep as 0.1\,K\,km\,s$^{-1}$ for CO integrated intensities, we still only trace $\sim50\%$ of pixels at those column densities. This $\sim5\times10^{20}$\,cm$^{-2}$ column density threshold corresponds to a visual extinction A$_{\rm V}$ of $\sim 0.5$\,mag, and is in good agreement to observational studies of CO-dark molecular gas \citep[e.g.][]{Klaassen2005,Grenier2005,Langer2014,Burton2015}. Along lines of sight that include a mixture of both low and high density regions, one must therefore be aware that CO will trace the highest density regions relatively well, whilst potentially being "blind" to more than 50\% of all the low density molecular material along the line of sight. 

These threshold column densities are also in agreement with \citet[][]{smith2014}, although they suggest that most of the CO-dark gas is in inter-arm molecular filaments. However, due to the effects of self-gravity and feedback in disturbing the smooth appearance of ISM structures, the morphology of the CO-dark gas here is significantly different to that of \citet[][]{smith2014}. Figure~\ref{fig:top-down} illustrates this more clearly, where we can see that most of the molecular gas which does not have associated CO (that we can see in green), is not in the form of smooth filamentary structures, but instead as more diffuse and disordered structures. The regions with high CO column density (in pink/white) are often associated with either peaks of density or even entire filamentary ridges in the inter-arm regions, reinforcing the notion that CO is only able to trace smaller portions of larger molecular complexes (see also Figure~\ref{fig:top-down-emission}). This is in good agreement with the idea that observed molecular clouds are like ``tips of icebergs'' \citep[e.g.][]{Pringle2001}.


\section{Extreme clouds and effect of environment}
\label{section:extreme_and_environment}

In this section we investigate how the properties of GMCs depend on galactic environment, by separating the sample of GMCs into arm and inter-arm clouds (in Sect.~\ref{section:arm_inter-arm}). We also investigate the environment of the more extreme clouds, and check how such clouds would be perceived from an observer's perspective (in Sect.~\ref{section:extreme}). 

\subsection{Arm $versus$ inter-arm regions}
\label{section:arm_inter-arm}

\begin{figure}
\centering
\includegraphics[width=0.39\textwidth]{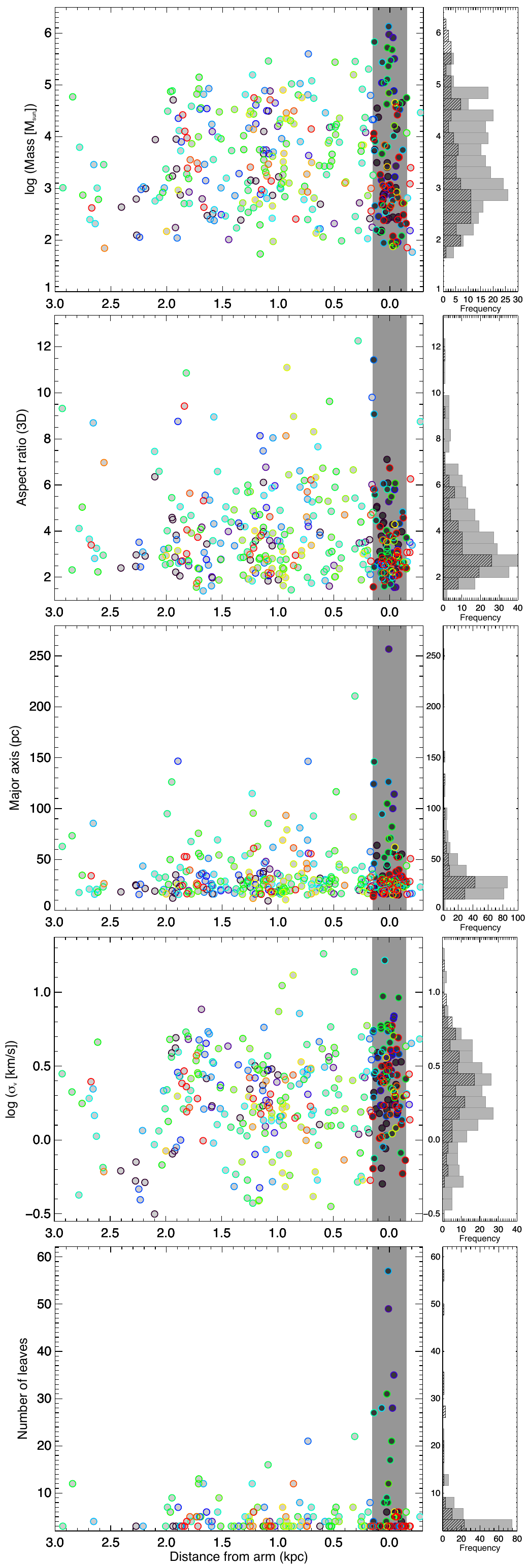}
\vspace{-0.3cm}
\caption{Distribution of mass (top), aspect ratio (second-row), major axis (third-row), line of sight velocity dispersion (fourth-row), and number of leaves (bottom), as a function of distance from the arm, for the GMCs from the PPP H$_{2}$ density. Black-filled circles (and grey-shaded area) show arm clouds, and grey-filled circles are inter-arm clouds. The colour-coding of the circles is as in Fig.~\ref{fig:top-down-molecular}, and is a proxy of the $z$ coordinate of the centre of each cloud. On the right of each panel are the distributions of these variables for arm clouds (dashed-black histograms), and for inter-arm clouds (grey-filled histograms).}
\label{fig:distance_aspect}
\end{figure}

\begin{table*}
\caption{Statistical properties of GMCs in the spiral arm and inter-arm regions, from the 3D PPP extraction using H$_{2}$ densities.}
\label{tab:stat_properties_arm_inter-arm}
\renewcommand{\footnoterule}{}  
\begin{tabular}{l | c c c c | c c c c c }
\hline 
\hline
\multirow{2}{2.0cm}{Property}	& \multicolumn{4}{c}{Arm Clouds}			& \,\,\,\,\, & \multicolumn{4}{c}{Inter-arm Clouds}		\\
						& Median$^{(a)}$ & Mean$^{(b)}$  & Skewness & Kurtosis & & Median$^{(a)}$ & Mean$^{(b)}$  & Skewness & Kurtosis \\
\hline 
\hline
$log($Mass (M$_{\odot}$)$)$		& $3.1\pm0.9$	& $3.5\pm1.1$	& 0.7 & -0.6	& & $3.4\pm0.7$	& $3.5\pm0.9$	& 0.2 & -0.9	\\
\hline
Aspect ratio (3D)				& $2.9\pm0.9$ 	& $3.5\pm1.6$  & 1.9 & 5.5 	& & $3.5\pm1.2$	& $4.0\pm2.0$	& 1.4	 & 2.0	\\
\hline
Major axis (pc) 				&  $27\pm11$ 	&  $38\pm34$	& 3.5	 & 16.1	& & $27\pm9$		& $35\pm25$	& 3.0	 & 13.4	\\
\hline
$\sigma_{v}$ (km\,s$^{-1}$)		& $2.5\pm1.1$ & $2.9\pm2.2$ 	& 2.7 & 12.0 	& & $1.9\pm0.9$	& $2.4\pm2.1$	& 3.4 & 18.5	\\
\hline
Number of leaves 				& $1.0\pm1.5$	& $5.1\pm9.6$	& 3.4	 & 12.3	& & $3.0\pm1.0$	& $1.0\pm1.0$	& 2.9 & 12.6	\\
\hline 
\end{tabular}
\flushleft
$^{(a)}$ Uncertainties derived as the mean value of the absolute deviation of the first and third quartiles from the median.\\
$^{(b)}$ Quoted uncertainty refers to the standard deviation.\\
\end{table*}

In order to understand whether the properties of clouds are affected by the different surrounding conditions, we divided the extracted GMCs into arm and inter-arm clouds, based on the projected distance from the arm in the $x-y$ plane. We adopted an arm width of 300\,pc for this simulation based on the surface density distribution from a top-down view. This is also similar to the arm width found for the Milky Way arms \citep[of $\sim$400\,pc, see][]{vallee2014b}. 
 
We used the clouds from the H$_{2}$ extraction, as it is more representative of the entire GMC complexes (see Sect.~\ref{section:ppp_ppv}), and we plot the distribution of the cloud properties as a function of the distance of clouds with respect to the spiral arm in Fig.~\ref{fig:distance_aspect}. We also show the histograms of the properties of the two sub-samples (arm clouds in dashed-black histograms, and inter-arm clouds in grey). The statistical properties of the two sub-samples of clouds are summarised in Table~\ref{tab:stat_properties_arm_inter-arm}. 

Although the differences between the properties of arm and inter-arm clouds are not particularly striking, there are some tendencies in the mean values and shapes of the distributions. In particular, even though the distributions of masses and major axis are similar (with a rather flat distribution with cloud masses ranging from $\sim10^2$ to $\sim10^{6}$\,M$_{\odot}$, and a relatively well peaked distribution of major axis around a value of 30-40\,pc), the largest and most massive clouds in the sample are in the arm, and they correspond to large GMC complexes (with masses larger than $10^{6}$\,M$_{\odot}$ and sizes larger that 100\,pc). Clouds in the arm also have a higher value of mean velocity dispersion and a higher mean number of leaves (i.e. are more sub-structured) compared to inter-arm clouds. This agrees with the observational results towards M51, e.g. from \citet[][]{Koda2009}, who suggested that the most massive GMCs are located exclusively in the spiral arms (with masses up to $10^{8}$\,M$_{\odot}$), and also from \citet[][]{Colombo2014} who found that clouds in the spiral arms have higher velocity dispersions than inter-arm clouds. On the other hand, although the mean aspect ratio of clouds is similar for arm and inter-arm regions, we find that the clouds reaching the highest aspect-ratio values in the sample (i.e. highly filamentary clouds) are all exclusively found in the inter-arm regions or in the process of entering the spiral arm. These clouds are reminiscent of the giant molecular filaments found in the Milky Way by \citet{Ragan2014}, and the filamentary GMCs found by \citet{Koda2009} in the inter-arms as a result of shear, both of which showing low velocity dispersions. These results highlight the fact that the arm is more ``chaotic'', and therefore there is a higher frequency of cloud-cloud collisions/interactions within the arms \citep[][]{DobbsPDC2015}, which increases the mean velocity dispersion in clouds, and results in more complex morphological structures. 

From the CO-density extraction we would recover similar trends for the arm and inter-arm clouds.  With the CO-emission clouds, however, these environmental trends are smoothed out, and become statistically hard to distinguish, although some of the most isolated filamentary clouds in the inter-arm are still singled out. However, as noted in Sect.~\ref{section:ppp_ppv}, with the severe blending inherited from our perspective and resolution, the properties of the clouds extracted in PPV space are not accurate tracers of the properties of the underlying clouds, and hence we cannot conclude whether we could potentially recover these statistical trends. To do so, we would need to repeat this study at a higher resolution (both in PPP and PPV) to minimise the projection limitations, which is beyond the scope of the current paper.

\subsection{Extreme clouds}
\label{section:extreme}

\begin{figure*}
\centering
\includegraphics[width=\textwidth]{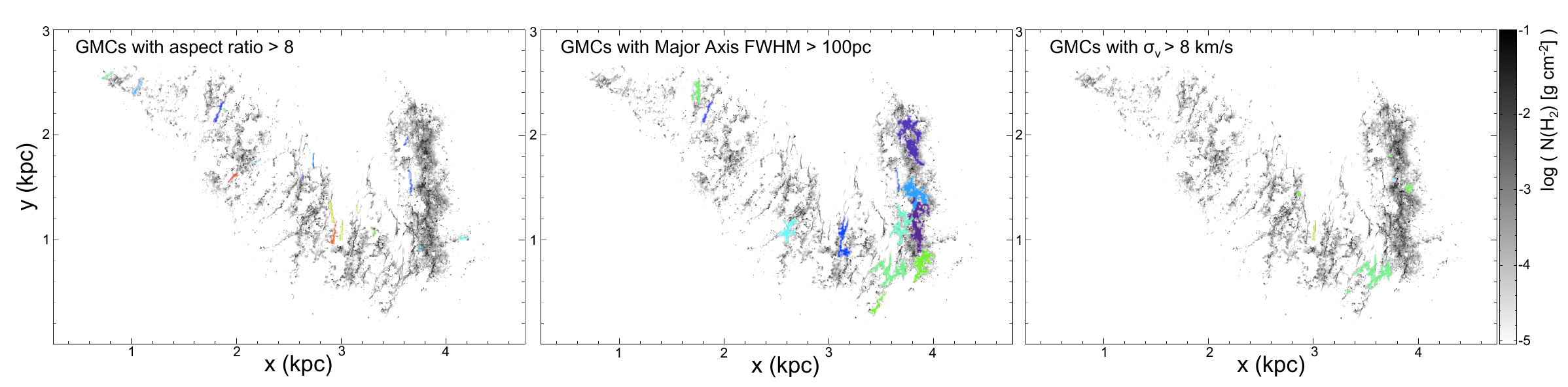}
\vspace{-0.4cm}
\caption{Top down view of the simulation, showing the position of the most extreme clouds from the H$_{2}$ density extraction, in terms of highest aspect ratio (of the left), major axis (centre) and velocity dispersion (right). The grey-scale shows the H$_{2}$ column densities, while the colours are the same as in Fig.\ref{fig:top-down-molecular}, i.e. each colour shows a different cloud, and is related to their $z$-coordinate (blue being below the plane, green in the plane, and red above the plane).}
\label{fig:extreme_clouds}
\end{figure*}

\begin{figure*}
\centering
\includegraphics[width=\textwidth]{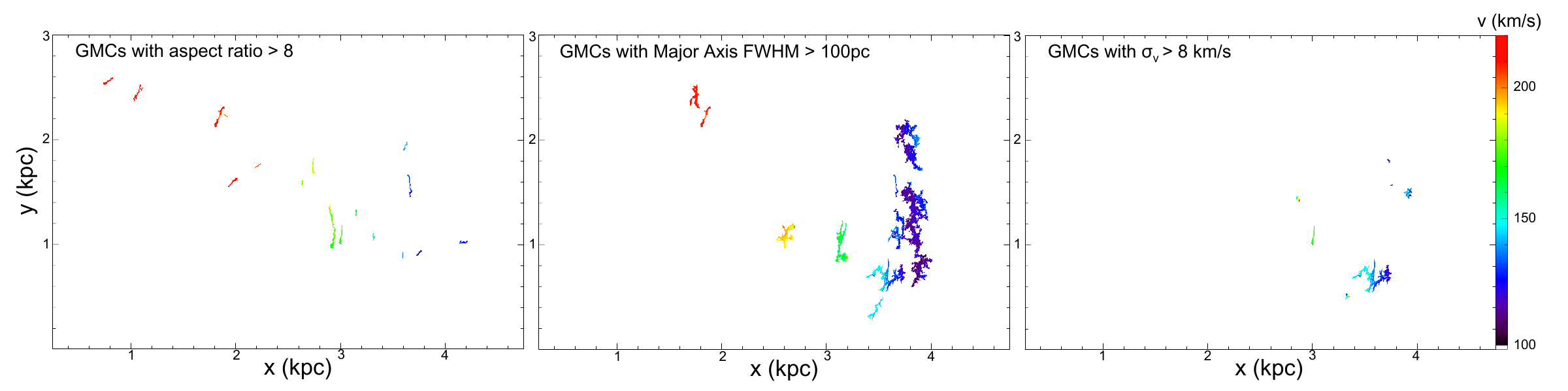}
\vspace{-0.4cm}
\caption{Line-of-sight velocity fields for the same clouds as in Fig.~\ref{fig:extreme_clouds} in colour scale, with velocities ranging from 100\,km\,s$^{-1}$ to 220\,km\,s$^{-1}$, for an observer positioned at top-left corner of the simulation, at $z=0$\,kpc.}
\label{fig:v_extreme_clouds}
\end{figure*}

In this section we pay particular attention to the clouds that form the tails of the distributions from Sect.~\ref{section:arm_inter-arm}. In particular, we investigated the highly filamentary clouds within the simulation by isolating the most elongated GMCs, with aspect ratios greater than 8. We also investigated the largest complexes, with major axis greater than 100\,pc, as well as the clouds with the highest velocity dispersions in the sample ($\sigma_{v}$ greater than 8\,km\,s$^{-1}$). The distribution of these clouds, as well as the velocity fields within them (as projected along the line-of-sight for our chosen observer position), are shown in Figs.~\ref{fig:extreme_clouds} and \ref{fig:v_extreme_clouds}. We discuss them in more detail in the following sections.

\subsubsection{Highly filamentary clouds}
\label{section:elongated_clouds}

Given the spatial resolution of our grid, selecting the most elongated GMCs from our sample as the clouds with an aspect ratio $>8$ guarantees that our clouds are at least $\sim$40\,pc in length. With this criteria we single out the very long and thin filamentary clouds, equivalent to the ``giant molecular filaments'' as observed in our Galaxy \citep[e.g. by][]{Jackson2010,Goodman2014,Ragan2014,Wang2015,Zucker2015}. We find a total of 18 clouds with aspect ratio greater than 8, i.e. about 5\% of all the clouds extracted for this section of the galaxy. If relaxing the threshold of an aspect ratio to 6, we would recover a total of 43 clouds (i.e. 12\% of all clouds), which are still filamentary but with lengths as low as $\sim$30\,pc.

\begin{figure}
\centering
\includegraphics[width=0.33\textwidth]{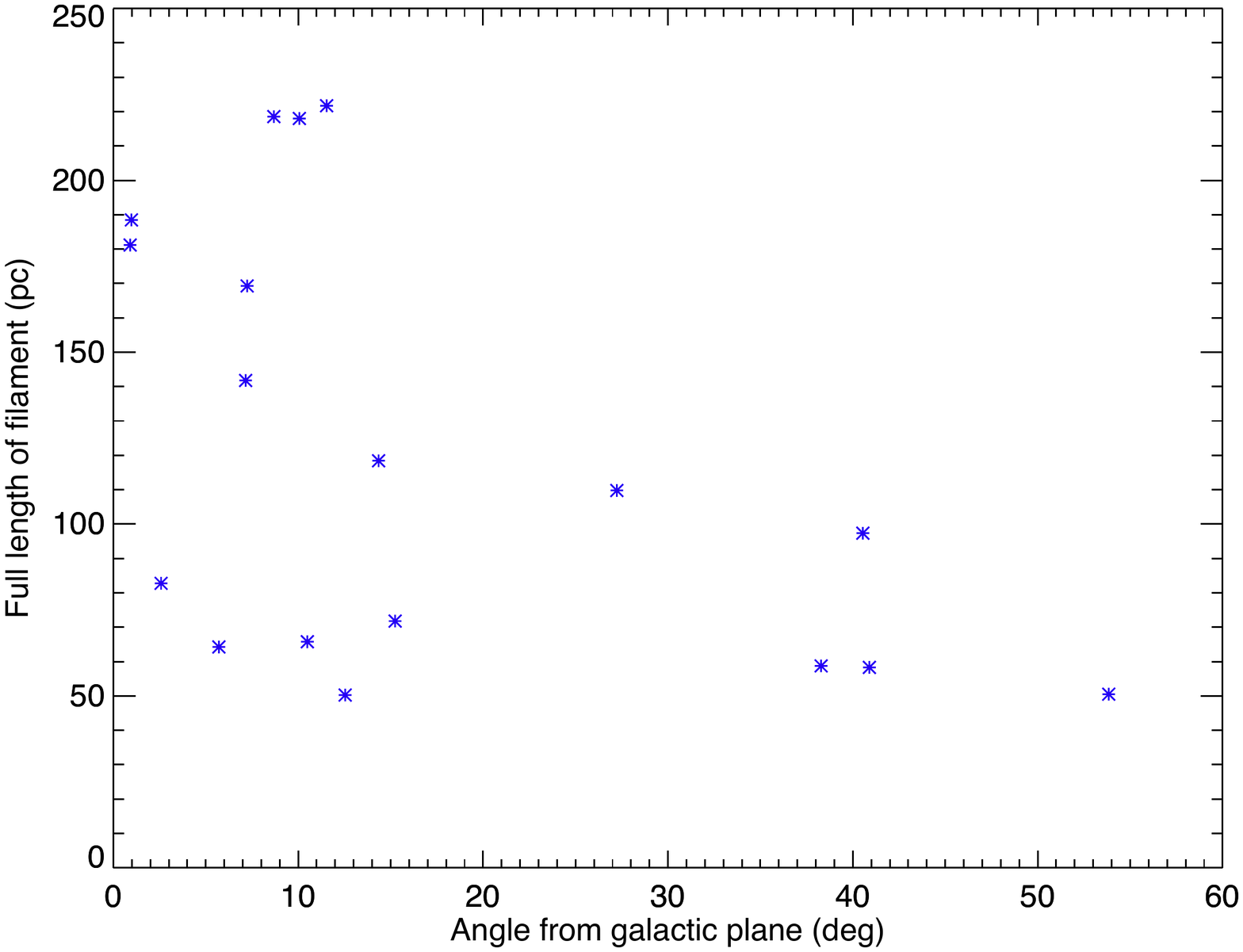}\\
\includegraphics[width=0.33\textwidth]{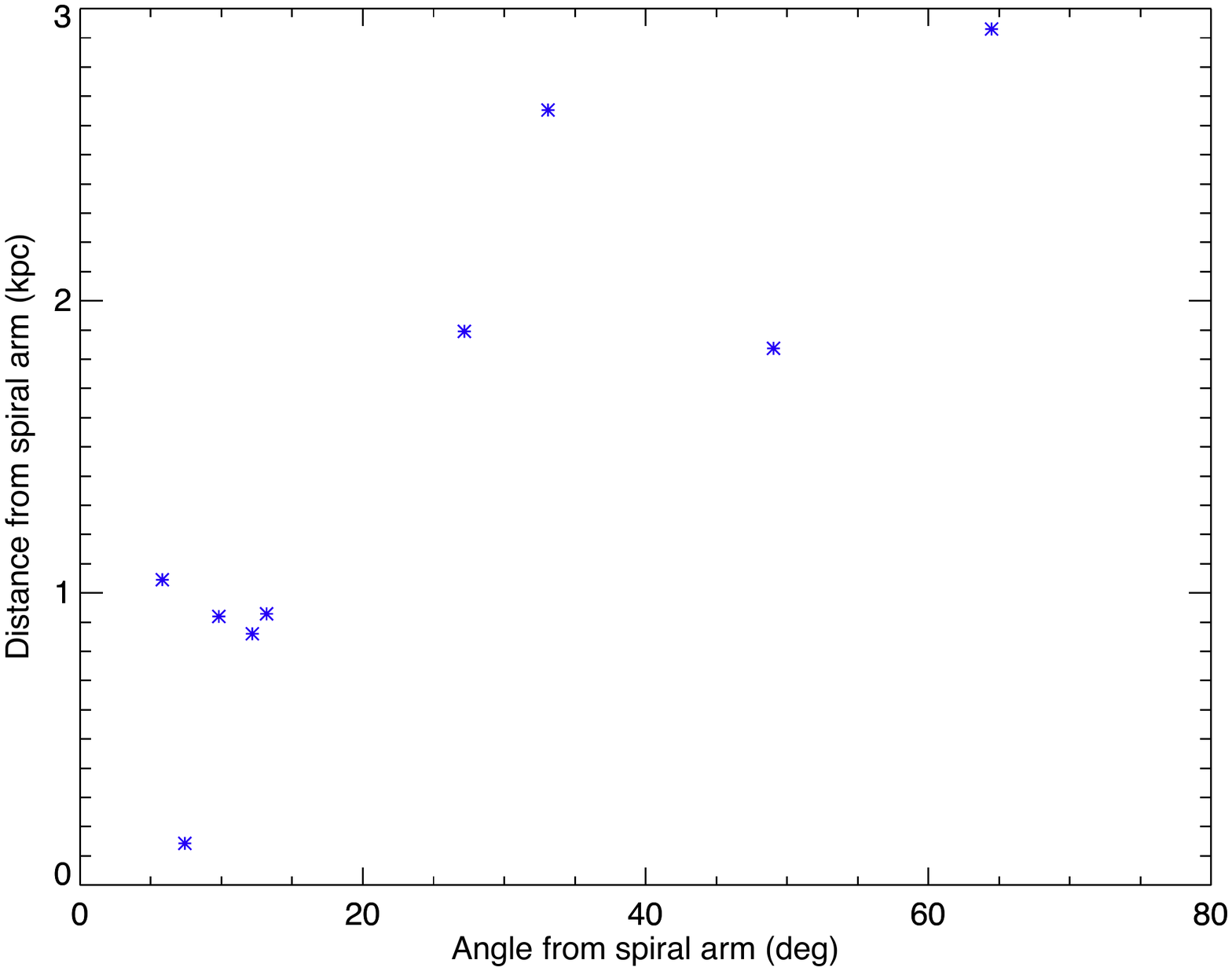}\\
\includegraphics[width=0.33\textwidth]{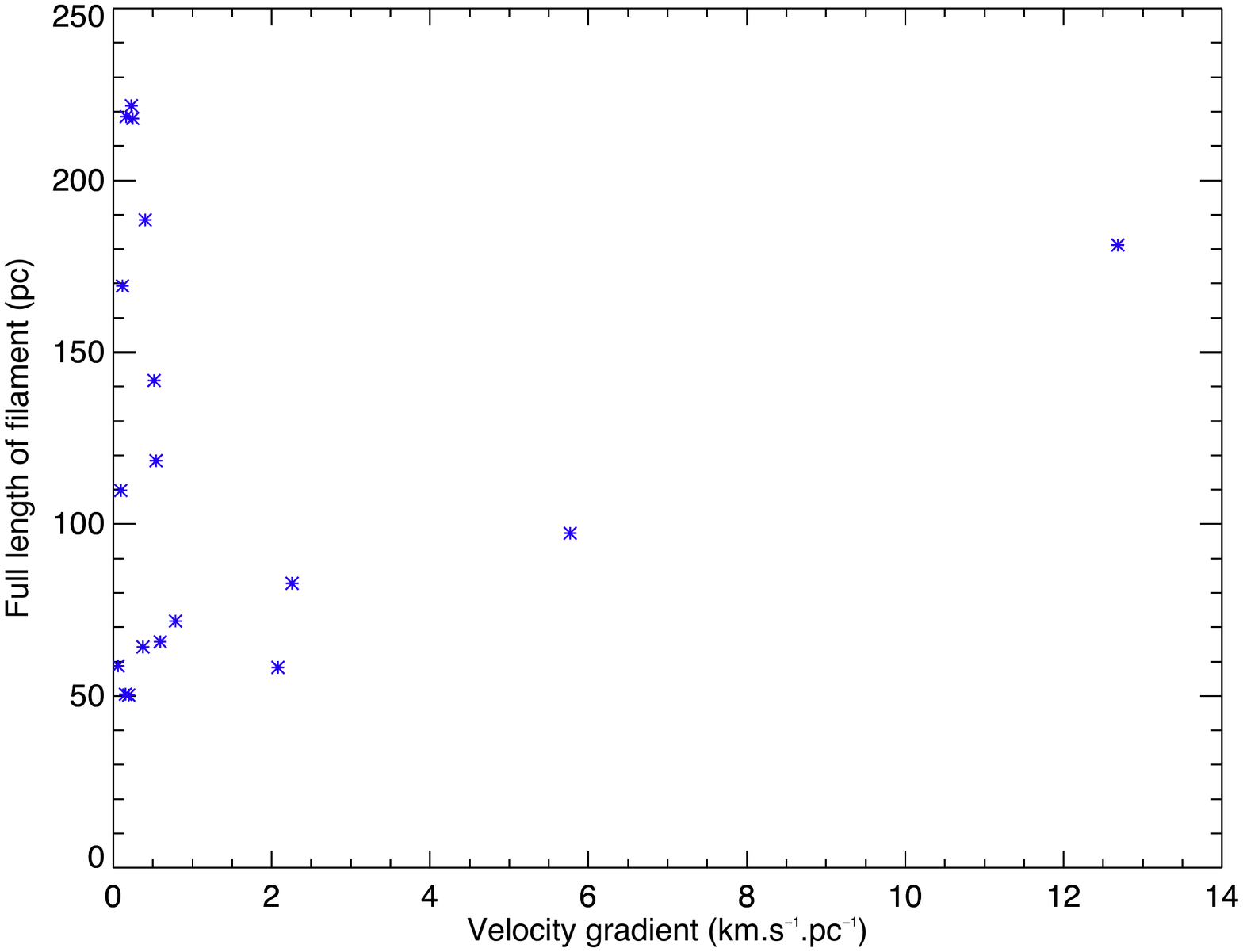}\\
\caption{{\it Top:} Full length of the filaments plotted as a function of their inclination with respect to the galactic plane (0$^{\circ}$ being on the plane); {\it Centre:} Distance of the filaments longer than 100\,pc from the spiral arm, and a function of their angle with respect to the spiral arm axis (0$^{\circ}$ being aligned with arm); and {\it Bottom:} Full length of the filaments plotted as a function of the velocity gradient across the filament.}
\label{fig:angle_plane}
\end{figure}

The position and velocities of the longer filaments (with aspect ratio $>8$) can be seen on the left panels of Figs.~\ref{fig:extreme_clouds} and \ref{fig:v_extreme_clouds}. We find that all of these long molecular filaments are situated either in the inter-arm region or in the process of joining the arm. Their radial profiles are unresolved in the 5\,pc grid we used for this study, meaning that their actual aspect ratio is higher than the values we derive. We have estimated the full length of these filaments 
by determining the maximum distance between any two points within each filament. We also estimated the inclination of the major axis of the filaments with respect to the galactic plane, and with respect to the spiral arm. The distribution of filament lengths as a function of the angle with the galactic plane is shown in the top panel of Fig.~\ref{fig:angle_plane}. The central panel of Fig.~\ref{fig:angle_plane} shows the distribution of the angles with respect to the spiral arm as a function of the cloud's distance to the spiral arm, but only for the longest filaments of our sample (i.e. those longer than 100\,pc).  We also estimated the velocity gradients across each cloud, by identifying the highest velocity difference between any two points belonging to each filament, and dividing by the separation of those two points. The velocity gradients of these filaments are shown in the lower panel of Fig.~\ref{fig:angle_plane} as a function of filament lengths. The H$_{2}$ masses of these filamentary clouds range from $2\times10^{2}$ to $3\times10^{4}$\,M$_{\odot}$. If considering only the longest filaments of the sample ($>$100\,pc), these have a mean mass of $10^{4}$\,M$_{\odot}$, and a mean mass per unit length of the order of $\sim80$\,M$_{\odot}$\,pc$^{-1}$. These masses per unit length are marginally smaller than those found in Galactic filaments by \citet[][]{Wang2015} or \citet[][]{Zucker2015}, but similar to those from \citet[][]{Ragan2014} when taking only the dense gas mass. Again, the smaller mass per unit length in our simulations could be a consequence of the relatively low maximum densities reached, due to the input of feedback.

We find that the longest filamentary clouds can stretch for nearly 250\,pc in full length, most of which are well aligned with the galactic plane, with inclination angles below 15$^{\circ}$. A few filaments have higher inclinations (up to $\sim$55$^{\circ}$), but we generally do not find long vertical filaments. We also do not find any particular correlation between the velocity gradients across filaments and their inclination with respect to the galactic plane. The range of angles with respect to the spiral arm is more variable, but the longest filaments clearly show a trend of increasing alignment with the spiral arm as they approach it. These preferred filament directions are a consequence of the differential rotation of the gas in the galaxy, and the encountering of the spiral arm potential. When clouds exit a spiral arm, they come out as spiral arm spurs, showing more chaotic directions (even perpendicular to the arm). As travelling through the inter-arm region, they then start to stretch and re-align due to the shear from the galactic rotation. By the time they are close to entering the spiral arm, the long filamentary clouds have major axes well aligned axis with the arm ($<$20$^{\circ}$). Unsurprisingly, smaller filaments do not show such a clear trend, as they are less affected by the galactic shear. 

The left panel of Fig.~\ref{fig:v_extreme_clouds} shows that these highly filamentary clouds have very uniform velocities, with smooth velocity gradients throughout. From the lower panel of Fig.~\ref{fig:angle_plane} we can see that these gradients are typically below 1\,km\,s$^{-1}$\,pc$^{-1}$, with only a couple of cases reaching more than 5\,km\,s$^{-1}$\,pc$^{-1}$. Only one filamentary cloud has a large velocity dispersion (cloud at $x\sim3$ and $y\sim1$\,kpc, common to the left and right panels of Figs.~\ref{fig:extreme_clouds} and \ref{fig:v_extreme_clouds}), and that is due to the very localised large velocity gradient ($\sim$13\,km\,s$^{-1}$\,pc$^{-1}$) towards its ``sourthern'' end - perhaps due to a recent SN feedback event. The remaining (pristine) part of the cloud is effectively smooth in terms of velocity.

These results are in agreement with the large inter-arm filaments found by \citet{Koda2009} in M51, and the giant molecular filaments found in the Galaxy \citep[e.g. by][]{Jackson2010,Ragan2014}, where velocity-coherent filaments with lengths ranging from $\sim$50\,pc up to $\sim$230\,pc, with velocity gradients much smaller than $\sim$1\,km\,s$^{-1}$\,pc$^{-1}$ (i.e. smaller than those we find\footnote{Note that \citet[][]{Ragan2014} estimate the maximum velocity difference within the filament, but divide that by the full length of the filament, not the actual separation between the two points with the highest velocity difference - this would bring their velocity gradients up to values closer to what we find here.}) have been reported in the inter-arm regions. \citet[][]{Goodman2014} suggested that the ``Nessie'' filament found by \citet[][]{Jackson2010} is in fact coincident with a spiral arm, and they find part of their support with an analogy to the structures found in the galaxy simulations by \citet[][]{smith2014}. However, those simulations, unlike the ones we present here, have no self-gravity or feedback included. As noted in \citet[][]{duarte-cabral2015}, both self-gravity and feedback are essential ingredients, particularly when trying to reproduce the distribution of the molecular gas in galaxy models. In the simulations we present here, where both gravity and feedback are included, we do not find such long filaments in the arms, as these quickly merge and interact with other clouds, collecting the gas into a clumpier medium making up large GMC complexes (a more detailed follow up of the time evolution of such clouds will be the subject of future work). Instead, we find long filaments aligned with the spiral arm just before they enter it, i.e. before they start interacting with arm clouds (e.g. see Fil-4 in Fig.~\ref{fig:top-down-emission}). The line-of-sight velocities at this entry point are quite close to those of the arm itself. Given the observational uncertainties when estimating kinematical distances, it could be that the Nessie filament is not in the arm itself, but indeed close to its entry point. 

To investigate how these giant filamentary clouds would be perceived from an observer's perspective, we examined the four giant filaments labeled Fil-1 to 4 in Fig.~\ref{fig:top-down-emission} as seen through the synthetic observations of the H$_{2}$ column densities and CO ($1-0$) emission (See Fig.~\ref{fig:four_filaments_ppv} in Appendix~\ref{app1}). We found that these filamentary clouds are highly unresolved in the coarser resolution maps, and they are hard to recognise in crowded emission areas due to severe line of sight confusion. At higher-resolution, the filamentary nature of clouds reappears, but the giant molecular filaments tend to be broken up into smaller filaments, some of which can be relatively faint in CO emission (see e.g. Fil-4). 

\subsubsection{Large GMC complexes}
\label{section:high_ma}

By selecting all clouds with a major axis greater than 100\,pc (irrespective of their aspect ratio), we retrieve 13 clouds ($\sim 4\%$ of all clouds) which are a mix of long filamentary clouds or large GMC complexes. The position and velocity field of these GMCs with large major axes can be seen in the central panels of Figs.~\ref{fig:extreme_clouds} and \ref{fig:v_extreme_clouds}. All the larger clouds in the inter-arm regions correspond to more filamentary clouds or a network of filaments (e.g. see cloud Cx-1 from Fig.~\ref{fig:top-down-emission}, and corresponding synthetic emission maps in Fig.~\ref{fig:complexes_ppv}).

In contrast, the large GMC complexes are found exclusively in the arm, as a result of the increased concentration of clouds and cloud-mergers in the arm. Unsurprisingly, these GMC complexes are also the most massive and sub-structured GMCs of the sample. Although these large complexes are coherent in density, they are not bound structures at these scales, and one could wonder of their ``physical'' significance as one large structure. Nevertheless, the concentration of material in the arm is such that clouds often form a ``continuum'' of material, bringing together a large mass reservoir surrounding the numerous over-densities at smaller scales inside these GMCs, possibly making the arms a preferred place for the formation of the most massive stars accreting from larger scales than their own parent clump.

\subsubsection{High velocity dispersion clouds}
\label{section:high_vdisp_cloud}

The other sub-set of clouds we selected were those with a large velocity dispersion (with $\sigma_{v}$ greater than 8\,km\,s$^{-1}$), which amounts to a total of 7 clouds ($\sim 2\%$ of all clouds). If following the Larson's size-linewidth relation, we would expect these clouds to correspond to large GMC complexes. However, the clouds with the highest velocity dispersions and large velocity gradients are typically quite small, where the gradients may arise from a localised and recent feedback event. The only exception is the large complex at $x\sim3.5$ and $y\sim0.8$\,kpc. This complex is particularly interesting, as it is one, large, coherent structure in density (picked up as one large cloud in both H$_{2}$ and CO densities), but it is experiencing a particularly complex velocity structure, with a gradient of $\sim$40\,km\,s$^{-1}$ across it. In fact, this cloud is the only in our sample that is experiencing a merger/collision at this particular timeframe, where a long filamentary cloud, nearly parallel to arm, has a velocity of $\sim$150\,km\,s$^{-1}$, and is being compressed against spiral arm clouds (at velocities of $\sim$110\,km\,s$^{-1}$). With this complex velocity structure, this cloud is in fact split into several CO emission clouds in PPV. 


\begin{figure}
\includegraphics[width=0.48\textwidth]{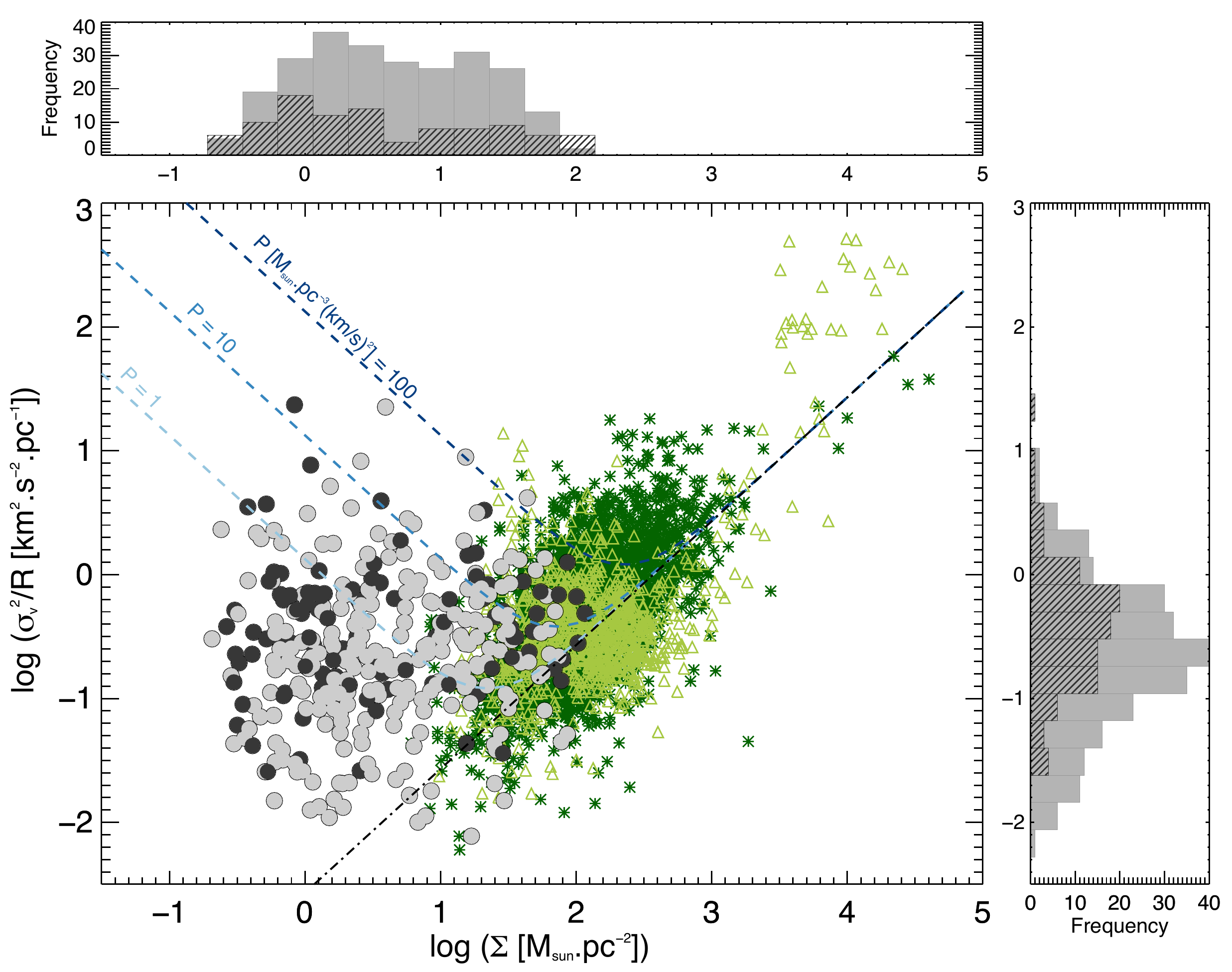}
\vspace{-0.5cm}
\caption{Characteristic size-linewidth coefficient ($\sigma_v^{2}/R$), as a function of gas surface density $\Sigma$ for all resolved GMCs extracted from the H$_{2}$ densities as circles (grey being inter-arm clouds, and black being arm clouds). The light-green triangles are a compilation of Galactic GMCs, while the dark-green star symbols are GMCs from extragalactic studies \citep[from][]{Rosolowsky2003,Ros07,Bolatto2008,Heyer2009,Rathborne2009,Santangelo2010,Wong2011,Swinbank2011,Ginsburg2012,Wei2012,Giannetti2013,Battersby2014,Colombo2014,Walker2015}. The dashed curves show the expected force balance (kinetic, gravitational and external pressure), for different values of external pressure, from $P = 1$ to $100$\,M$_{\odot}$\,pc$^{-3}$\,km$^{2}$\,s$^{-2}$ (which corresponds to $P/k \sim 5\times10^3 - 5\times10^5$ K\,cm$^{-3}$). The top histogram shows the distribution of H$_{2}$ surface densities for the arm and inter-arm GMCs from the simulations (in dashed-black and filled-grey histograms respectively). The histogram on the right-hand side shows the distribution of $\sigma_v^{2}/R$ for the simulated GMCs, with same colour-coding as the top histogram.}
\label{fig:correlation_plots}
\end{figure}

\section{Global equilibrium state of GMCs}
\label{section:equilibrium}

With observations of nearby star forming clouds, \citet{Larson1981} identified a number of scaling relations that described the global behaviour of a number of cloud properties. One such relation, indicates a proportionality between the velocity dispersion line-widths ($\sigma_v$),  and the radius for clouds as $R^{\Gamma}$ \citep[originally with $\Gamma \sim 0.38$, but later modified to 0.5, e.g.][]{Solomon1987}. However, this relation has been revised over the years, with an increasing number of surveys available at higher sensitivities, able to detect both Galactic and extragalactic GMCs over a wider range of conditions \citep[e.g.][]{Heyer2009,Leroy2015}. Indeed, \citet{Heyer2009} pointed to a correlation between the gas surface densities $\Sigma$ and the size-linewidth coefficient, represented here as $\sigma_v^{2}/R$, shown in Fig.~\ref{fig:correlation_plots} for a compilation of Galactic and Extragalactic GMCs, as well as the clouds presented here. 

\begin{figure}
\includegraphics[width=0.48\textwidth]{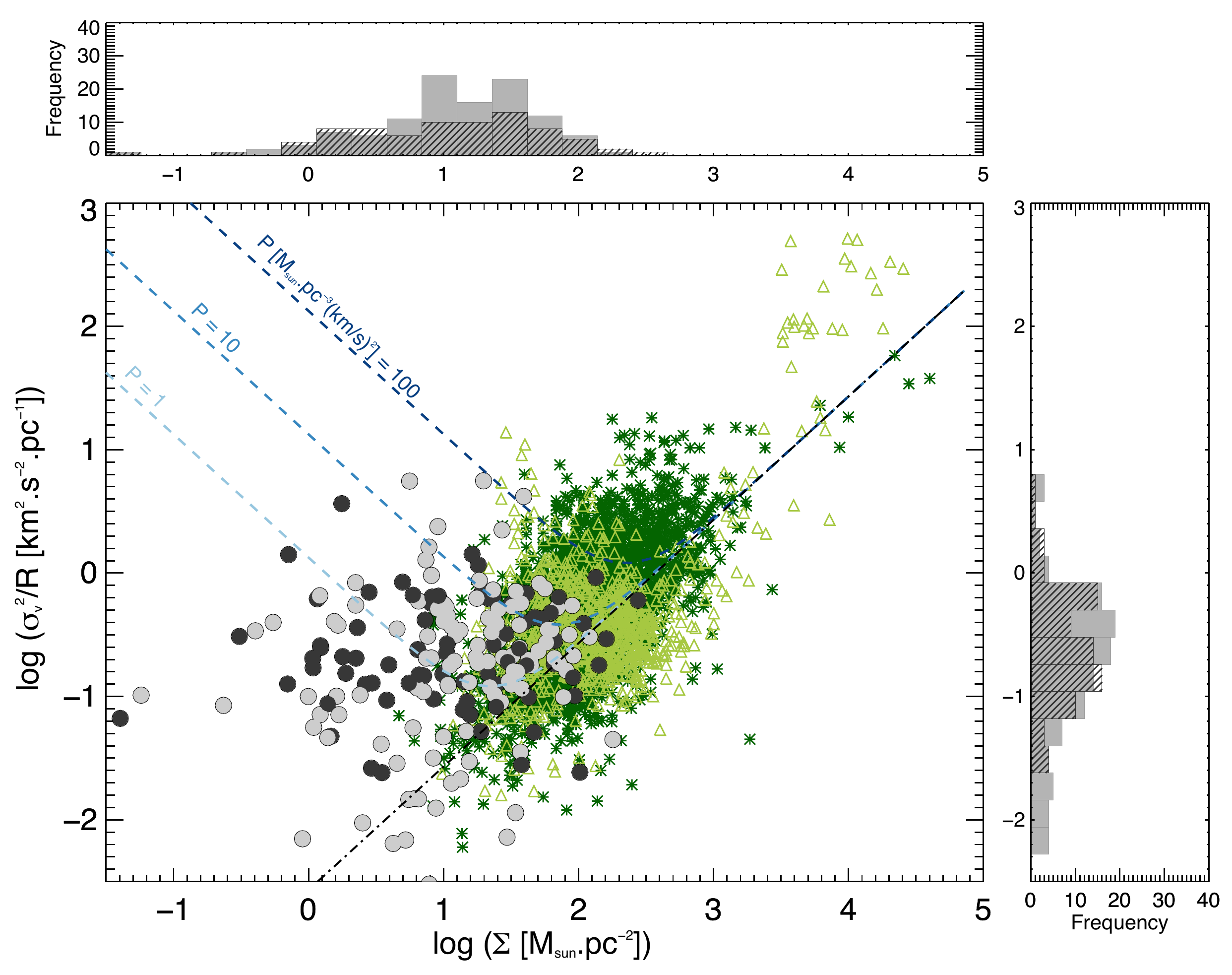}
\vspace{-0.4cm}
\caption{Characteristic size-linewidth coefficient ($\sigma_v^{2}/R$), as a function of gas surface density ($\Sigma$), as in Fig.~\ref{fig:correlation_plots}, but now with the clouds from our CO-density extraction, which focuses mostly on the denser H$_{2}$ clouds (or simply the denser parts of large H$_{2}$ GMCs) of the sample, hence the average surface densities are now higher. Symbols as in Fig.~\ref{fig:correlation_plots}.}
\label{fig:correlation_plots_CO}
\end{figure}

This proportionality has been interpreted by \citet{Ball2011} as evidence of gravitationally bound clouds with a similar virial parameter (defined as the balance between gravitational and kinetic energy, i.e. $\alpha_{vir} = 2E_k/E_g \propto \sigma_v^2/(\Sigma\,R)$). This relation can also be re-interpreted as a consequence of the force balance between turbulence, pressure and gravity (e.g. Traficante et al. in prep.). For low surface densities, there is a non-negligible turbulent pressure from the medium which can dominate over gravity in driving the kinematics (pressure-confined regime). At higher gas surface densities, the gravity becomes dominant, and can potentially be enough to drive all of the $\sigma_v$, in which case the gravitational force ($\propto \Sigma$) drives the increase of the kinetic force ($\propto  \sigma_v^2/R$). The coloured dashed lines in Fig.~\ref{fig:correlation_plots} show this for three values of external pressure, with $\alpha_{vir} =1$ shown as a black dash-dotted line. The theoretical and observational values of the external pressure generated by the neutral ISM in the Galaxy are of the order of $P/k \sim 10^{4} - 10^{6}$\,K\,cm$^{-3}$ \citep[e.g.][]{Elmegreen1989,Bertoldi1992,Sakamoto2011}, i.e. of the order of the pressures plotted in Fig.~\ref{fig:correlation_plots}. The turn-over into a regime of gravity-driven kinetic force, requires surface densities of the order of $\sim$100\,M$_{\odot}$\,pc$^{-2}$ (i.e. $\sim$0.02\,g\,cm$^{-2}$), for a $P/k$ of the order of $5\times10^{4}$\,K\,cm$^{-3}$.

The clouds from the galaxy simulation studied here could potentially be pressure confined by a relatively low-pressure medium, and there is no noticeable difference between arm and inter-arm clouds. The median value of the virial parameter, $\alpha_{vir}$, for the H$_{2}$ GMCs is of $\sim$20, suggestive of highly unbound clouds in the classical equilibrium analysis \citep[see also][]{Dobbs11}. Due to the input of feedback in the model at relatively low volume densities ($\sim$500\,cm$^{-3}$), not much material is actually allowed to reach the higher end of the surface densities, hence there are no simulated clouds populating the right-hand side of Fig.~\ref{fig:correlation_plots}. Nevertheless, irrespective of the higher density areas inside the GMCs not being captured, the average surface densities at the larger scales of the GMCs are reliable. The little observational equivalent of the low surface density clouds may be due to the observational limitations, such that observations typically capture only the denser parts of GMCs as smaller molecular clouds or clumps, where the gravitational forces dominate and control the kinematic acceleration. In fact, if we do the same exercise with the CO-density extracted clouds (Fig.~\ref{fig:correlation_plots_CO}), which trace the denser GMCs (or simply the denser parts of the larger GMCs), we do obtain a smaller median virial parameter of $\alpha_{vir} \sim5$, and clouds move closer to the turn-over point between pressure and gravity-dominated regimes. Since we have a high density threshold at which to input feedback, we cannot test how the clouds would move further into the high-surface density regime with this simulation.


\section{Summary and conclusions}
\label{section:conclusions}

In this paper, we have investigated the GMC population from an SPH simulation of a portion of a two-armed spiral galaxy, that includes cooling and heating of the ISM, H$_{2}$ and CO chemistry, self-gravity and SNe feedback. We investigated the morphological and basic dynamical properties of GMCs both from the 3-dimensional densities of H$_{2}$ and CO, as well as from an observer's perspective using synthetic observations of the CO emission, and we further linked these properties to the clouds' context in the galaxy. The main results from this study can be summarised as follows:
\begin{itemize}
\item The global ratio between atomic and molecular hydrogen in this simulation is close to the observed value in the Milky Way, as is the CO-to-H$_{2}$ relation. This resolves the problem of underproducing molecular material which was an issue in the lower-resolution simulations of galaxies, as found by \citet[][]{duarte-cabral2015}. Nevertheless, the feedback in this simulation is still input at relatively low densities, which means that we cannot probe the density regimes where the gas is fully molecular (our GMCs are $\lesssim$30\% molecular).
\item The statistical properties of clouds as traced by the 3D distribution of CO and those from the actual distribution of H$_{2}$ gas are similar. However, we find that CO densities are a good tracer of the very high density H$_{2}$ gas, but they only trace a smaller fraction of the total molecular material. This has three main consequences: {\it i)} CO clouds miss most of the low-density H$_{2}$ GMCs where there is little CO; {\it ii)} The regions where there is enough CO can often be unresolved density peaks, leading to missing the smaller (even if dense) H$_{2}$ clouds; and {\it iii)} The largest H$_{2}$ GMC complexes of the sample are broken up into smaller less massive clouds in CO, as CO densities miss the low density material that connects the different substructures within the GMCs.
\item When taking the perspective of a Galactic observer, we find that the spatial resolution plays a particularly critical role in order not to blend clouds that are not physically connected. With high-spatial resolution and high-sensitivity observations, this projection effect is significantly less severe, and the only remaining caveat is the fact that the CO emission is merely sensitive to the peaks of the CO densities, and cannot trace the full extent of the underlying molecular gas. This is a consequence of the extremely weak CO emission for low CO column densities, and the resulting increasing amount of H$_{2}$ low-density gas which has no CO-emission associated. This will impact the shapes of the statistical distributions of cloud properties, particularly for the distribution of aspect ratios, and masses. Nevertheless, with a combined study of the H$_{2}$ column densities in the plane of the sky, and the CO emission, one should be able to ``connect the dots'' and recover the link between the different CO-emission clouds and the larger scale GMCs.
\item From the extraction of clouds from the H$_{2}$ densities, we find that clouds in the spiral arms have higher velocity dispersions, and that the arm hosts the largest, most massive, and more sub-structured clouds of the sample. We interpret these results as a consequence of the higher concentration of material along spiral arms, which promotes a larger number of interactions/mergers that provides a framework to build up larger GMC complexes.
\item From the same H$_{2}$ density-based extraction, we find that the highly filamentary clouds of our sample reach lengths as large as $\sim$250\,pc , whilst their widths are unresolved (i.e smaller than 10\,pc in width). They are exclusively found in the inter-arm region, or else as clouds in the process of entering the spiral arm. Most of these long filaments have low inclinations with respect to the galactic plane, and they are increasingly aligned with the spiral arms as they approach them. All of our highly filamentary clouds have smooth velocity gradients throughout, as true analogues of the ``giant molecular filaments'' found in the Milky Way.
\item If in equilibrium, our sample of GMCs would be mainly pressure confined, rather than gravitationally dominated.
This pressure confinement may only be dominant at the scales at which we probe most of our clouds, with overall low surface densities. If probing the denser regions within the GMCs we would expect the clouds to move into a regime where gravitational forces become dominant.
\end{itemize}

Ideally, to track the effects of the journey of clouds in the galaxy, we would need to follow clouds over time - but time is something that is not accessible through observations. In order to mimic this observational limitation, here we have studied the large complexes of gas within one single time-frame of the simulations. By studying a large sample of clouds at any one moment, the statistics should reflect the different time- (and space-)dependent properties of clouds. Although our results suggest that the global ``median'' properties of the GMCs from the simulation do not seem to show strong differences between arm and inter-arm environments, the environment does seem to have an impact on dictating the properties of some clouds. In particular, our results would suggest that large filamentary clouds found in the inter-arm regions are not formed locally, but are instead the result of galactic-shear-induced re-shaping of the molecular gas that was once part of a spiral arm. This highlights the fact that molecular clouds are not ``formed'' and ``destroyed'' in a simplistic scenario of atomic-to-molecular-to-atomic transition. Instead the dense molecular gas is generally long-lived, reshaped and exchanged, during its galactic journey \citep[as also suggested by][]{Scoville1979,Koda2009,DobbsPDC2015}. In future work, we will follow the evolution clouds over time \citep[e.g. as in][]{Dobbs2013}, and investigate how the statistical differences of cloud properties with environment correspond to the actual time evolution of clouds as they travel through the galaxy. 

\section*{Acknowledgments}

We thank the anonymous referee for useful comments that helped strengthen the paper. ADC and CLD acknowledge funding from the European Research Council for the FP7 ERC starting grant project LOCALSTAR. This work used the DiRAC Complexity system, operated by the University of Leicester IT Services, which forms part of the STFC DiRAC HPC Facility (www.dirac.ac.uk ). This equipment is funded by BIS National E-Infrastructure capital grant ST/K000373/1 and  STFC DiRAC Operations grant ST/K0003259/1. DiRAC is part of the National E-Infrastructure. This work also used the University of Exeter Supercomputer, a DiRAC Facility jointly funded by STFC, the Large Facilities Capital Fund of BIS, and the University of Exeter. Figure~\ref{fig:sim} was produced using SPLASH \citep{Price2007}. This work has used the CAMELOT project \citep[][DOI: (in progress); http://camelot-project.org/]{Ginsburg2016}.

\bibliographystyle{mnras}
\bibliography{fb} 


\appendix

\section{Synthetic observations of select clouds}
\label{app1}

\begin{figure*}
\includegraphics[width=0.8\textwidth]{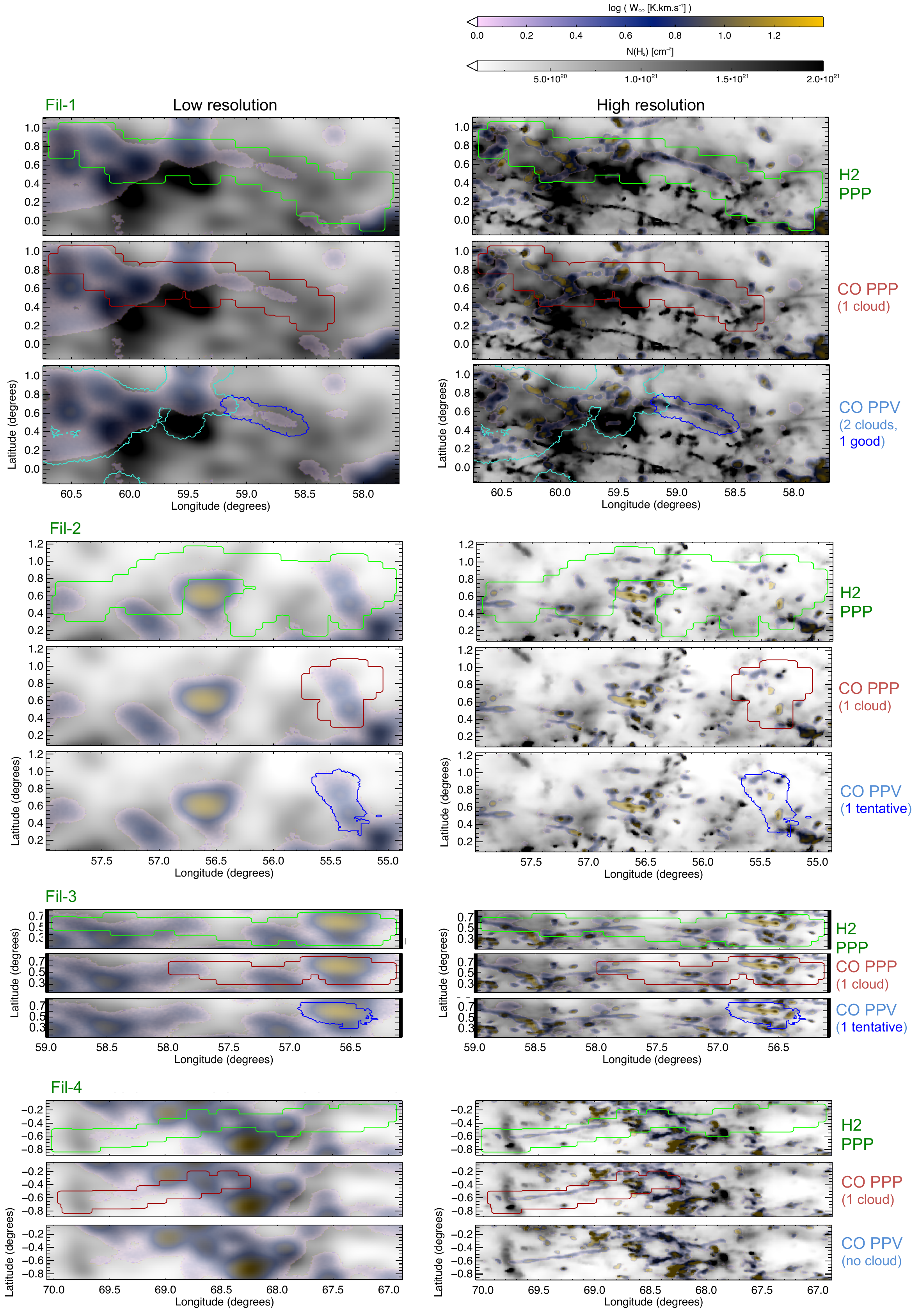}
\caption{$Longitude-latitude$ plot of the synthetic observations of the H$_{2}$ column density (in gray scale), and the CO ($1-0$) integrated intensity (in colour-scale with transparency) for the four filamentary cloud labeled in Fig.~\ref{fig:top-down-emission} (Fil-1 to 4). The top-row of each set of panels shows the H$_{2}$-PPP cloud (in green contours); the second row shows the corresponding CO-PPP cloud (red contours); and the bottom row shows the overlapping CO-PPV clouds, with the best-match CO-PPV cloud shown in dark blue, and other satellite clouds that overlap partially with the original cloud in turquoise contours. The left column maps have an angular resolution of 0.2$^{\circ}$ ($\sim$10\,pc), and the right column a higher resolution of 36''.}
\label{fig:four_filaments_ppv}
\end{figure*}

\begin{figure*}
\includegraphics[width=0.8\textwidth]{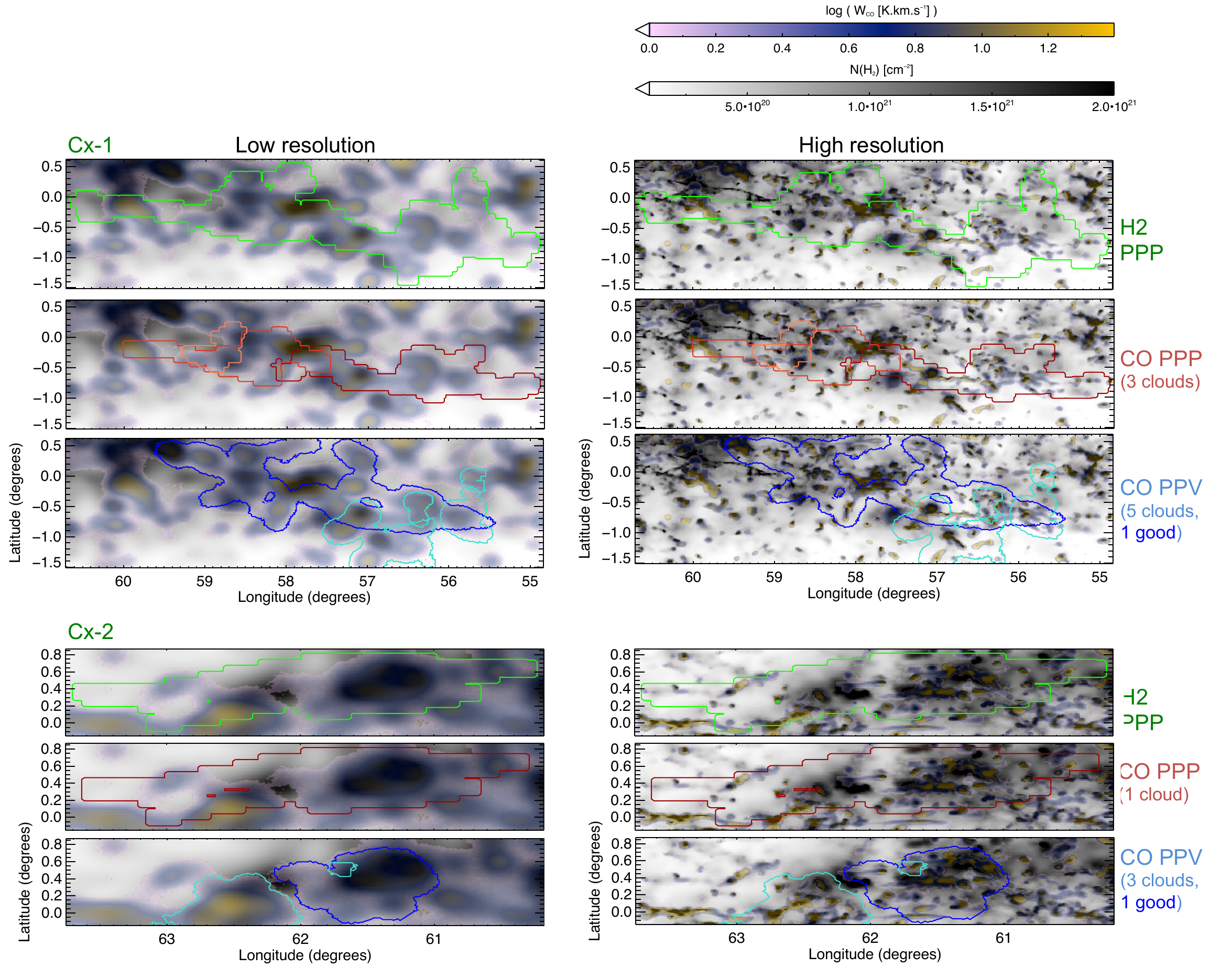}
\caption{Same as Fig.~\ref{fig:four_filaments_ppv} for the molecular cloud complexes labeled as Cx-1 and Cx-2 from Fig.~\ref{fig:top-down-emission}.}
\label{fig:complexes_ppv}
\end{figure*}

To illustrate how clouds in PPP space are seen and recovered from a PPV perspective as a Galactic observer, we have taken a sample of 6 clouds, as per Fig.~\ref{fig:top-down-emission}, and plotted the projected extent of their H$_{2}$-PPP masks, their corresponding CO-PPP masks, and any CO-PPV cloud whose mask overlaps with the original cloud in PPV space. This is shown in Figs.\,\ref{fig:four_filaments_ppv} and \ref{fig:complexes_ppv}. In these figures, the cloud masks are overlaid on the map of the projected total H$_{2}$ column density, and CO integrated intensity (integrated around the velocity ranges of each cloud). For each cloud, we show the observer perspective at two resolutions: the 0.2$^{\circ}$ resolution on the left (i.e. from where we extracted the CO-PPV clouds), and at higher resolution on the right, with a beam size of 36''.

From Figs.\,\ref{fig:four_filaments_ppv} and \ref{fig:complexes_ppv} we can see that while some clouds have a good cloud match from the CO-PPV extraction, the exact coverage of these is not the same as the original cloud, due to severe blending along the line of sight (e.g. Cx-1), and/or the lack of CO in the lower density parts of the cloud (e.g. Fil-3). Among the better matches to portions of the original clouds are those of Fil-1, and Cx-2, while the two other clouds (Fil-4 and Fil-3) have no clear match in CO-PPV. For Fil-4 this is simply because the filament is too weak and too unresolved in the CO-PPV cube where we did the extraction, and hence indistinguishable in the lower resolution map. For Fil-3, even though the CO-PPV cloud coincides with the peak of the CO-density cloud, it is not clear whether this strong emission peak truly belongs to that particular density peak, as the distance we derived for it is not placing the cloud at the right position in PPP (i.e. our distances place this cloud on another, stronger, CO-density peak along the same line of sight, with the same velocity). The lack of emission as seen from the top-down perspective (Fig.~\ref{fig:top-down-emission}) would indeed suggest that we ought not to expect a CO-PPV counterpart for this cloud.

With high-sensitivity and high-resolution data, however, as shown in the right-hand panels, we can see that the blending of clouds would become less severe, and although we would only be able to pick up the high-density peaks within bigger molecular cloud complexes, we would at least be more confident that the derived properties for those clouds/clumps were not strongly affected by projection effects. This higher resolution is comparable to recent Galactic plane surveys, such as SEDIGISM (Schuller et al. in prep), CHIMPS \citep{Rigby2016}, COHRS \citep{Dempsey2013}, among others, and therefore we believe that the statistical properties of clouds derived from these surveys should be reliable, although bearing in mind that they are not isolated clouds, but part of larger complexes, undetectable in CO.

\bsp	
\label{lastpage}
\end{document}